\documentclass[%
 reprint,
 amsmath,amssymb,
 aps,
]{revtex4-2}

\usepackage{graphicx}
\usepackage{dcolumn}
\usepackage{bm}

\begin{document}


\title{Variational Self-Consistent Theory for Beam-Loaded Cavities}

\author{Adham Naji}\email{anaji@stanford.edu}
 \altaffiliation[]{}
\author{Sami Tantawi}\email{tantawi@slac.stanford.edu}%
 
\affiliation{%
 Particle Physics and Astrophysics Department, Stanford University, Stanford, CA 94305 \\
 SLAC National Linear Accelerator Laboratory, Stanford University, Menlo Park, CA 94025
}%

\date{\today}

\begin{abstract}
A new variational theory is presented for beam loading in microwave cavities. The beam--field interaction is formulated as a dynamical interaction whose stationarity according to Hamilton's principle will naturally lead to steady-state solutions that indicate how a cavity's resonant frequency, $Q$ and optimal coupling coefficient will detune as a result of the beam loading. A driven cavity Lagrangian is derived from first principles, including the effects of cavity wall losses, input power and beam interaction. The general formulation is applied to a typical klystron input cavity to predict the appropriate detuning parameters required to maximize the gain (or modulation depth) in the average Lorentz factor boost, $\langle \Delta\gamma \rangle$. Numerical examples are presented, showing agreement with the general detuning trends previously observed in the literature. The developed formulation carries several advantages for beam-loaded cavity structures. It provides a self-consistent model for the dynamical (nonlinear) beam--field interaction, a procedure for maximizing gain under beam-loading conditions, and a useful set of parameters to guide cavity-shape optimization during the design of beam-loaded systems. Enhanced clarity of the physical picture underlying the problem seems to be gained using this approach, allowing straightforward inclusion or exclusion of different field configurations in the calculation and expressing the final results in terms of measurable quantities. Two field configurations are discussed for the klystron input cavity, using finite magnetic confinement or no confinement at all.  Formulating the problem in a language that is directly accessible to the powerful techniques found in Hamiltonian dynamics and canonical transformations may potentially carry an additional advantage in terms of analytical computational gains, under suitable conditions. 
\end{abstract}

\maketitle



\section{\label{Intro}Introduction}

Beam loading of microwave cavities is an important phenomenon that is often seen in high-power vacuum tubes when a cavity's electromagnetic field interacts dynamically with an accelerated beam of electrons. As the beam gets accelerated or decelerated by the Lorentz forces imposed by the cavity's field, it acts back on the cavity, modifying its field. This dynamical interaction settles into a steady-state that exhibits observable shifts in both the cavity's resonant frequency and its quality factor ($Q$). Particle accelerators, satellites and radar systems that employ high-power vacuum electronics are example applications where such a phenomenon is observed and must be addressed for proper system design \cite{Gilmor,Antonsen2002}. For example, the detuning experienced when the beam is turned on in a particle accelerator may cause a mismatch for a klystron input cavity and a reduction in gain, as the coupling coefficient and resonant frequency no longer correspond to their original design values. Another typical example can be seen in high-power signalling applications, where a detuning in the radio frequency (rf) cavity may  lead to the emission of undesirable intermodulation distortion (IMD) products that must be suppressed \cite{Gilmor,Antonsen2002,Antonsen2004,Antonsen2005}. 

The literature on the theory of beam loading for azimuthally-symmetric cavities spans more than seven decades and includes a number of key studies; e.g.~\cite{Pierce,Branch,Antonsen2002,Antonsen2004,Antonsen2005}. Branch's classical treatment \cite{Branch} has had a strong analytical influence on many of the more recent efforts to investigate and develop beam-loading theory \cite{Antonsen2002,Antonsen2004}. The effect of beam loading is conventionally modelled in terms of circuit theory as a shunt admittance, $Y_{B}=G_{B}+iB_{B}$, that loads that rf cavity (representing the beam's influence), where $G_{B}$ denotes the beam-loading conductance and $B_{B}$ denotes the beam-loading susceptance. When the beam is turned on (``hot" operation), the ``cold" frequency $f_{\text{cold}}$ and quality factor $Q_{\text{cold}}$ are decreased according to the following approximation \cite{Antonsen2002}
\begin{eqnarray}
    Q_{\text{hot}}&=& \frac{Q_{\text{cold}}}{1+\left(G_{B}/G\right)},  \label{eq:Qhot}\\  \frac{f_{\text{hot}}-f_{\text{cold}}}{f_{\text{cold}}}&=&\frac{\Delta f}{f_{\text{cold}}}=\frac{-1}{2Q_{\text{cold}}}\frac{B_{B}}{G}, \label{eq:DeltaF}
\end{eqnarray}
where $G$ is the cavity's original (cold) shunt conductance.

To estimate $G_{B}/G$ and $B_{B}/G$ in cavity gaps, Branch's original treatment \cite{Branch} covered a premise where operation assumed certain idealized conditions, such as setting the axial magnetic flux for beam confinement to extreme values ($B_{z}$ set to zero or infinity) and ignoring space charge effects (ballistic analysis) \cite{Branch}. Important subsequent studies have attempted to develop the theory further and to include more realistic conditions using Particle-in-Cell (PIC) simulations, such as the effect of a finite $B_{z}$ flux strength, \cite{Antonsen2002,Antonsen2004}, and the effect of ac space charge, ~\cite{Antonsen2002,Antonsen2005}, compared to ballistic analysis. These studies have concluded that, for operation at moderate perveance levels (typically less than 3 $\mu$pervs \cite{Antonsen2005}), a first approximation that assumes infinite magnetic flux and uses ballistic beam-loading predictions would deviate only weakly from those that include finite axial magnetic flux \cite{Antonsen2004} or ac space charge effects \cite{Antonsen2005,Antonsen2002}. Deviation due to the latter effect is particularly small (less than 1\% in $\Delta f$) for common operation condition (below 3~$\mu$pervs). Furthermore, one of the striking observations found through PIC simulations in \cite{Antonsen2002} was that beam-loading is mainly influenced by perveance (not by beam current or voltage varying separately) and that $\Delta f$ generally tends to vary almost linearly with perveance. 

Methods of analysis that were used in these studies \cite{Pierce,Branch,Antonsen2002, Antonsen2004,Antonsen2005} have typically followed a combination of circuit analysis and Fourier field integrals to describe the beam--field interaction within a gap, where the maximum amplitude of the rf field was treated as a fixed quantity and the loading effect was modelled by the lumped admittance $Y_{B}$.  In this paper we follow a different formulation, motivated by how the dynamical interaction between the rf field and the beam seem to be naturally well-suited for variational analysis. We use a time-harmonic variational formulation to determine the stationary operating point upon which this dynamical interaction will settle, indicating the detuned frequency and $Q$ values for the cavity. We then apply the analysis to a typical klystron input cavity and discuss the effect of detuning on maximizing gain. For brevity, we present two cases in detail (ignoring space charge effects): one for an unconfined beam and one for a beam with finite axial confinement (focusing magnetic flux, $B_{z}$).  The beam-loading trends predicted by this analysis are shown, through numerical examples, to agree with the general trends previously reported in the literature. The developed theory characterizes a given cavity structure under given beam-loading conditions by two detuning parameters. These parameter can help us optimize cavity shapes and maximize gain under beam loading.    

As will be shown in the following sections, the physics underlying beam-loading phenomena and the addition or removal of different field configurations become relatively straightforward and systematic under the presented formulation. Another advantage of the presented formulation is how it develops a theoretical framework that connects directly to the powerful analytical tools found in Hamiltonian dynamics and canonical transformations \cite{Litchenberg,Percival,Goldstein,hori,Lathem, CARY}. Indeed, by leveraging advanced canonical transformations and Lie perturbation techniques under certain conditions, one can potentially render the system computations fully or partially analytic without having to explicitly solve the equations of motion~\cite{Litchenberg,hori,Lathem,CARY}.

The paper is organized as follows. We start in Section~\ref{Prelim} by discussing the field representation for the structure under consideration and by defining key parameters and relations in a form suitable for our variational formulation. In Section~\ref{Lagrangian} we proceed to formulate the Lagrangian for a general beam-loaded cavity with a single port. We apply the developed theory to the important class of cavities exemplified by a klystron input cavity in Section~\ref{InputCavity}, where we extract the two detuning parameters, denoted $\bar{X}$ and $\bar{Y}$, and discuss gain maximization. The theory is then applied to numerical examples in Section~\ref{Examples}. We conclude in Section~\ref{Conclusions} and give additional derivations in Appendices~\ref{Appx1}, \ref{Appx2} and \ref{Appx3}.


\section{\label{Prelim}Preliminary Definitions and Formulae}
\begin{figure}
\includegraphics[width=7.3cm]{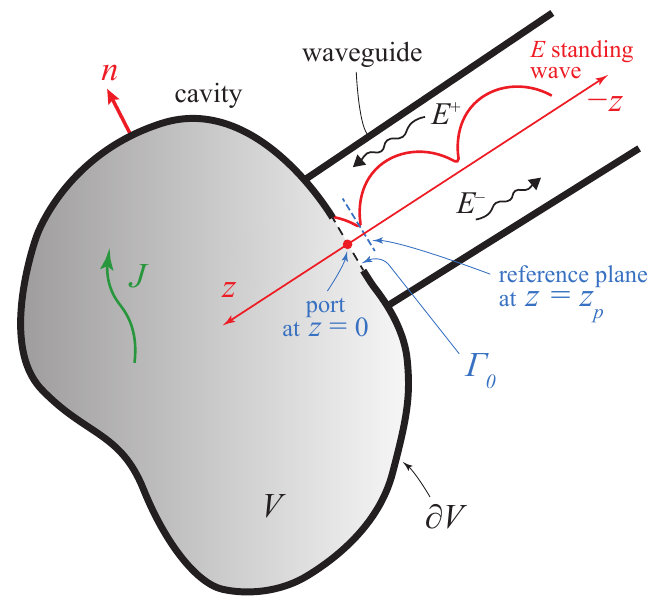}
\caption{\label{fig:cavity} An illustration of a single-port cavity with an arbitrary shape. The port connects to a feeding transmission line (e.g.~waveguide) and the walls are highly conductive. A current density $\bm{J}$ may exist inside the cavity. The surface normal vector points outwards and the port's reference plane, $z=z_{p}$, is chosen for convenience at the $E$-field standing wave's minimum (dip) nearest to the cavity walls.}
\end{figure}

In this section we establish the basic representation of the problem, define parameters and derive formulae in a form that will be useful for direct application in the Lagrangian formulation developed in Section~\ref{Lagrangian}.

\subsection{\label{geometry} Cavity and port field representations }
Consider a single-port cavity of an arbitrary, simply-connected shape, as shown in Fig~\ref{fig:cavity}. The domain of solution is the cavity's volume, $V\subset \mathbb{R}^{3}$, which is enclosed by a surface $\partial V$. The latter is made out of a good conductor and is geometrically closed, except for a small port that couples the cavity to a feeding transmission line. 

For time-harmonic fields with $e^{-i\omega t}$ time dependence,  Maxwell's curl equations,
$\nabla\times\bm{H}=-\epsilon\omega\bm{E}+\bm{J}$ and $   \nabla\times\bm{E}=i\omega\mu\bm{H}$, will lead to the driven wave equation
\begin{equation}
     \nabla^{2}\bm{E}+k^{2}\bm{E}=-i\omega\mu\bm{J}+\nabla(\nabla\cdot\bm{E})=-i\omega\mu\bm{J}, \label{eq:WaveEquation}   
\end{equation}
where $\omega$ is the operating (observed) angular frequency, $k=\omega\sqrt{\epsilon\mu}$ is the wavenumber, $\bm{E}$ is the electric field strength, $\bm{H}$ is the magnetic field strength, $\bm{J}$ is a current density present inside the cavity volume, $\epsilon$ is the permittivity, $\mu$ is the permeability, and with boldface fonts representing vectors. In writing Maxwell's equations and (\ref{eq:WaveEquation}) we assumed that the medium may be approximated as linear, isotropic and homogeneous, with $(\nabla\cdot\bm{E})\cong 0$.  The finite conductivity ($\sigma$) of the cavity walls is characterized by the surface impedance $\eta=(1-i)\xi$, where $\xi=1/(\sigma\delta_{s})$ is the surface resistance and $\delta_{s}=\sqrt{2/\omega\mu\sigma}$ is the skin depth \cite{schwinger,jackson,pozar}. The impedance boundary condition on the walls and the complex power flow into the conductor's skin can be written, respectively, as
\begin{equation}
    \hat{\bm{n}}\times\bm{E}=(1-i)\xi\bm{H} \ \ \ \ (\textrm{on }\partial V), \label{eq:BConWalls}
\end{equation}
\begin{eqnarray}
    \frac{1}{2}\int\limits_{\partial V}da (\bm{E}\times\bm{H}){\cdot} \hat{\bm{n}}{=}\frac{\eta}{2}\int\limits_{\partial V}da \bm{H}{\cdot}\bm{H}{=}\frac{1-i}{2}\xi\int\limits_{\partial V}da H^{2}, \label{eq:skinIntegral}
\end{eqnarray}
where $da$ denotes an element of surface area and $\hat{\bm{n}}$ is the outward-pointing unit normal vector of the surface.

As an idealized setup, if we momentarily take the same cavity with $\bm{J}{=}0$, perfectly conducting walls and no port, then solving the undriven wave equation ($\nabla^{2}\bm{E}+k^{2}\bm{E}=0$) within the cavity's volume and imposing the perfect boundary conditions will give us the cavity's eigenmodes $(\bm{e}_{j},\bm{h}_{j})$ and the corresponding eigenvalues $\omega_{j}$ (resonant frequencies). Here, the subscript $j$ is a positive integer index and the eigenmodes satisfy the curl equations
\begin{eqnarray}
    \nabla\times\bm{e}_{j}&=&i\omega_{j}\mu\bm{h}_{j}, \label{eq:curl_ei}\\
    \nabla\times\bm{h}_{j}&=&-i\omega_{j}\epsilon\bm{e}_{j}. \label{eq:curl_hi}
\end{eqnarray}

It is well-known that, within the solution space for the fields that can exist inside the cavity, one can always expand any field as a linear combination of the eigenmodes $(\bm{e}_{j},\bm{h}_{j})$ \cite{jackson,slaterbook,schwinger}, such as $\bm{H}=\sum_{j}\alpha_{j}\bm{h}_{j}$, where $\alpha_{j}=|\alpha_{j}|e^{i\phi_{\alpha_{j}}}$ are the eigenspectrum complex amplitudes; with a similar expansion for $\bm{E}$.

It follows from Rayleigh's principle \cite{Meirovitch,jeffreys,schwinger} that the eigenvalues of such a system are stationary points of the Rayleigh quotient which happens to represent the same system. If we are operating near the $i$-th mode, that mode's eigenfunctions $(\bm{e}_{i},\bm{h}_{i})$ will have the highest contribution (amplitude) in the eigenspectrum, compared to all other eigenmodes with $j\neq i$. If the operating field happens to coincide exactly with one of the $i$-th eigenmode, then the amplitudes $\alpha_{j}$ for other modes ($j\neq i$) in the expansion will be identically zero. This logic can be now extended to the case of perturbing our cavity from its idealized picture (undriven and lossless) back to its realizable setup, where the walls are made of good (but finite) conductivity, a beam current density $\bm{J}$ may exist inside the volume $V$ and a port may exist at the walls. Assuming that the $i$-th mode (typically the dominant mode) was the mode of operation before perturbation and that the perturbation is typically small, it is reasonable to consider that the added features have disturbed the cavity slightly, shifting the apparent resonant frequency $\omega$ off its spectral point $\omega_{i}$, and projecting the field distribution into the set of all possible eigenmodes, as in the sum $\sum_{j}\alpha_{j}\bm{h}_{j}$. Since it is expected that $|\alpha_{i}|$ will now be less that 1 but still much larger than the other expansion factors ($|\alpha_{i}|\gg |\alpha_{j\neq i}|$), one can choose to truncate the expansion at any number of terms in the neighborhood of the $i$-th (driven) mode. For the analysis of a beam-loaded cavity that is typically tuned to a specific resonant frequency with a high $Q$ factor, we assume it sufficient to use a single term (the $i$-th mode) representation to approximate the field. For example, the field $\bm{H}$, which we shall later use as the main variational variable in the formulation adopted in Section~\ref{Lagrangian}, can be written as,
\begin{eqnarray}
    \bm{H}\cong \alpha_{i}\bm{h}_{i}. \label{eq:EandH}
\end{eqnarray}
Consequently, the perturbation in the cavity, manifested through the frequency shift $\Delta\omega/\omega_{i}=(\omega-\omega_{i})/\omega_{i}$ when the cavity is loaded, is now being interpreted to be caused by an adjustment to the complex amplitude of the $i$-th mode that was perturbed by some function, $f(x,y,z)$, which happened to be proportional to the eigenmode itself. In variational language this can be expressed as, e.g., $\delta H=\varepsilon h_{i} \Rightarrow H\equiv h_{i}\pm\varepsilon h_{i}=(1\pm\varepsilon)h_{i}=\alpha_{i}h_{i}$, where $\varepsilon$ is a small paramater. In Section~\ref{Lagrangian} we solve the corresponding variational problem to find the stationary value of $\alpha_{i}$ for the beam-loaded cavity. 
 
In passing, it is worth noting that a perturbation of this type, affecting the complex amplitude of the eigenfunction in a single-mode expansion, may appear at first as a trivial way to proceed in the analysis since the amplitude may seem as a simple scalar that would not affect the shape of the eigenfunction or the wave equation itself. This would have been the case, indeed, had the wave equation remained undriven or had the Lagrangian, $\mathcal{L}$, of the system (as will be discussed in Section~\ref{Lagrangian}) remained homogeneous in $\bm{E}^{2}$ or $\bm{H}^{2}$, \cite{jeffreys,schwinger}. This is, however, not the case, since the port coupling and the beam current that act to perturb the cavity, as a driven system, have different functional forms that will not change linearly with the rest of the terms in $\mathcal{L}$ as we tune $\alpha_{i}$. This subtle point is of central importance for the variational formulation developed in Sections~\ref{Lagrangian} and \ref{InputCavity}.

The transmission line connected to the port is typically a waveguide operating above cutoff to facilitate rf power transfer into the cavity. Inputted power, $P_{\text{in}}$, is provided by an external source and is incident upon the port along the $+\hat{\bm{z}}$ direction, whereas any emitted power from the cavity, $P_{e}$, flows in the $\hat{\bm{n}}=-\hat{\bm{z}}$ direction, as shown in Fig~\ref{fig:cavity}. The fields propagating along the waveguide are taken to be proportional to the guide's dominant eigenmode $(\bm{e}_{p},\bm{h}_{p})$, and we can express the incident fields due to input power (denoted $\bm{E}^{+},\bm{H}^{+}$) and the cavity's emitted fields (denoted $\bm{E}_{e},\bm{H}_{e}$) as
\begin{eqnarray}
    (\bm{E}_{e},\bm{H}_{e})&=&a_{e}(\bm{e}_{p},\bm{h}_{p}), \label{eq:Ee,He}\\
    (\bm{E}^{+},\bm{H}^{+})&=&a^{+}(\bm{e}_{p},-\bm{h}_{p}), \label{eq:E+,H+}
\end{eqnarray}
where $a^{+}=|a^{+}|e^{i\phi^{+}}$ and $a_{e}=|a_{e}|e^{i\phi_{\text{em}}}$ are complex amplitudes. The propagating field functions $(\bm{e}_{p},\bm{h}_{p})$ are real functions for the waveguide above cutoff.

Since an incident wave ($\bm{E}^{+}, \bm{H}^{+}$) will generally reflect at the port, giving rise to ($\bm{E}^{-}, \bm{H}^{-}$), a standing wave will form inside the guide. For convenience we choose the arbitrary port plane to be where the incident and reflected electric fields interfere destructively closest to the cavity walls, giving a dip in the electric field's standing wave (see Fig~\ref{fig:cavity}). In practice, the typical port size is electrically small and the incident wave undergoes almost a total reflection, giving a point of minimum electric field and maximum magnetic field (approximately $E_{p}\rightarrow 0$ and $H_{p}\rightarrow 2H^{+}$) close to the cavity's wall (approximately $z_{p}\rightarrow 0$). When the cavity is emitting a field $(\bm{E}_{e},\bm{H}_{e})$, we can now see that  the total field observed at $z=z_{p}$ will be simply written as $(\bm{E}_{e},\bm{H}_{e}+2\bm{H}^{+})$.

\subsection{\label{Q_definitions} Parameter definitions and connecting formulae}
For the $(\bm{e}_{i},\bm{h}_{i})$ mode at resonance, $\omega=\omega_{i}$, the stand-off between electric and magnetic fields within the ideal cavity, as implied by Poynting's theorem \cite{slaterbook,pozar,jackson}, will result in equal electric and magnetic stored energies, $u_{i}$. Let us denote the phases of the complex functions $\bm{e}_{i}$ and $\bm{h}_{i}$ by $\phi_{e}$ and $\phi_{h}$, respectively. As will be seen in the following sections, it will be convenient for our formulation to take $\phi_{h}= 0$ as a reference value, which immediately implies that $\phi_{e}=\pi/2$. Since for an ideal cavity there is no power flux out of the cavity's surface, we must have a vanishing integral $\oint_{\partial V}da\ \hat{\bm{n}}\cdot(\bm{e}_{i}\times\bm{h}_{i})$. Using Divergence theorem, the vector identity $\nabla\cdot(\bm{e}_{i}\times\bm{h}_{i})=(\nabla\times\bm{e}_{i})\cdot\bm{h}_{i}-(\nabla\times\bm{h}_{i})\cdot\bm{e}_{i}$, (\ref{eq:curl_ei}) and (\ref{eq:curl_hi}) we define $u_{i}$ for the ideal cavity as
\begin{eqnarray}
    0=\frac{1}{2}\int\limits_{V}dV\nabla{\cdot}(\bm{e}_{i}\times\bm{h}_{i})&=&\frac{i\omega_{i}}{2}\int\limits_{V}dV (\mu h^{2}_{i}+\epsilon e^{2}_{i}) \nonumber \\
    \Rightarrow  \frac{1}{2}\int\limits_{V}dV \mu h^{2}_{i}&=&-\frac{1}{2}\int\limits_{V}dV\epsilon e^{2}_{i}\equiv u_{i}, \label{eq:ui}
\end{eqnarray}
where $dV$ denotes the element of volume and the notation $h_{i}^{2}$ means $\bm{h}_{i}\cdot\bm{h}_{i}$ (this should not be confused with the magnitude $|\bm{h}_{i}|^{2}$, which carries no phases). 

Any added dissipation due to imperfect walls in practice will change the cavity's $Q$ factor and perturb its resonant frequency, $\omega\neq\omega_{i}$, as the observed fields $(\bm{E},\bm{H})$ and their stored energy $U_{i}$ deviate slightly from their values for the ideal eigenmode $(\bm{e}_{i},\bm{h}_{i})$ \cite{slaterbook,schwinger}. With the help of (\ref{eq:curl_ei})--(\ref{eq:EandH}) and (\ref{eq:ui}), the intrinsic quality factor $Q_{i}=\omega_{i}U_{i}/P_{l_{i}}$, where $P_{l_{i}}$ is the power lost to the walls, may be written as (see Appendix~\ref{Appx1})
\begin{eqnarray}
    Q_{i}&\cong&\omega_{i}\frac{\mu\int_{V}dV h_{i}^{2}}{\xi\int_{\partial V}da\ h_{i}^{2}}=-\omega_{i}\frac{\epsilon\int_{V}dV e_{i}^{2}}{\xi\int_{\partial V}da\ h_{i}^{2}}= \omega_{i}\frac{u_{i}}{p_{i}}, \label{eq:Qi}
\end{eqnarray}
where we have defined the scaled quantities as
\begin{equation}
U_{i}=|\alpha_{i}|^{2}u_{i}, \ \ \ \ P_{l_{i}}=|\alpha_{i}|^{2}p_{i}. \label{eq:Ui_uiANDPl_pi}
\end{equation}

In a similar fashion, as we equip the cavity with a port, the external quality factor $Q_{e}=\omega_{i}U_{i}/P_{e}$, where $P_{e}$ is the power emitted from the cavity through the exiting field $(\bm{E}_{e},\bm{H}_{e})$ defined in (\ref{eq:Ee,He}), is written as (see Appendix~\ref{Appx1})
\begin{equation}
    Q_{e}\cong\frac{-\omega_{i}e^{-2i\Delta\phi}\alpha^{2}_{i}\int_{V}dV \epsilon e_{i}^{2}}{a^{2}_{e}\int_{p}da \ (\bm{e}_{p}\times\bm{h}_{p})\cdot\hat{\bm{n}}}    =\frac{\omega_{i}e^{-2i\Delta\phi}\alpha^{2}_{i}\int_{V}dV \mu h_{i}^{2}}{a^{2}_{e}\int_{p}da \ (\bm{e}_{p}\times\bm{h}_{p})\cdot\hat{\bm{n}}}, \label{eq:QeTempEq}
\end{equation}
where the sign $\int_{p}$ denotes integration over the cross section of the port, $\Delta\phi=\phi_{H}-\phi_{\text{em}}$ and the phase of $\bm{H}$ is
denoted $\phi_{H}$, which is equal to $\phi_{\alpha_{i}}+\phi_{h}=\phi_{\alpha_{i}}$ since we have set $\phi_{h}=0$ as a reference.  For an electrically small port with a port plane that is close enough to the cavity's walls ($z_{p}\rightarrow 0$), one can approximate the phase $\phi_{\text{em}}$ of the emitted field as $\phi_{\text{em}}\rightarrow \phi_{\alpha_{i}}$, $\Delta\phi\rightarrow 0$. Indeed, this is expected, since the magnetic field $\bm{H}$ inside the cavity, with phase equal to $\phi_{\alpha_{i}}$, will act to impress an equivalent current moment across the port, which will drive the emitted fields through the guide with phase $\phi_{\text{em}}\rightarrow \phi_{\alpha_{i}}$.  Equation (\ref{eq:QeTempEq}) is now reduced to
\begin{equation}
    Q_{e}=\frac{\omega_{i}\alpha^{2}_{i}\int_{V}dV \mu h_{i}^{2}}{a^{2}_{e}\int_{p}da \ (\bm{e}_{p}\times\bm{h}_{p})\cdot\hat{\bm{n}}}=\frac{-\omega_{i}\alpha^{2}_{i}\int_{V}dV \epsilon e_{i}^{2}}{a^{2}_{e}\int_{p}da \ (\bm{e}_{p}\times\bm{h}_{p})\cdot\hat{\bm{n}}}. \label{eq:Qefinal}
\end{equation}

The coupling coefficient, $\beta_{c}$, and total quality factor, $Q_{t}$, are defined and related to $Q_{i}$ and $Q_{e}$ by
\begin{equation}
\beta_{c}=\frac{Q_{i}}{Q_{e}}, \ \ \frac{1}{Q_{t}}=\frac{1}{Q_{i}}+\frac{1}{Q_{e}}, \ \  Q_{e}=Q_{t}\frac{1+\beta_{c}}{\beta_{c}}   \label{eq:QtinTermsOfQi}
\end{equation}

The power inputted through the port, $P_{\text{in}}$ is given by (see Appendix~\ref{Appx1}) 
\begin{eqnarray}
P_{\text{in}}&{=}&-\frac{1}{2}\int\limits_{p}da \bm{E}^{+}{\times}\bm{H}^{+*}{\cdot}\hat{\bm{n}}=e^{-2i\phi^{+}} \frac{a^{+2}\alpha_{i}^{2}}{a_{e}^{2}}\frac{\omega_{i}u_{i}}{Q_{e}},\label{eq:Pin} 
\end{eqnarray}

From the above definitions, one can also extract the following relations, which will be needed in Section~\ref{Lagrangian},
\begin{equation}
    \frac{|a^{+}|}{|a_{e}|}=\frac{1}{|\alpha_{i}|}\sqrt{\frac{P_{\text{in}}Q_{e}}{\omega_{i}u_{i}}}=\frac{e^{i\phi_{\alpha_{i}}}}{\alpha_{i}}\sqrt{\frac{P_{\text{in}}Q_{e}}{\omega_{i}u_{i}}}. \label{eq:useful1}
\end{equation}
\begin{eqnarray}
&&\int\limits_{p}dV\bm{E}_{e}\times\bm{H}^{+}\cdot\hat{\bm{n}}=-a^{+}a_{e}\int\limits_{p}da\ \bm{e}_{p}{\times}\bm{h}_{p}\cdot\hat{\bm{n}} \nonumber \\
&&=-2\frac{a^{+} \alpha_{i}^{2}\omega_{i}u_{i}}{a_{e}Q_{e}}=-2\alpha_{i}e^{i\phi^{+}}\sqrt{\frac{P_{\text{in}}\omega_{i}u_{i}}{Q_{e}}}, \label{eq:useful2}
\end{eqnarray}
\begin{equation}
    \xi\int\limits_{\partial V}da\ h_{i}^{2}=\frac{\omega_{i}}{Q_{i}}\mu\int\limits_{V}dV h_{i}^{2}, \label{eq:useful3}
\end{equation}
\begin{equation}
    \frac{1}{2}\int\limits_{p}da\ \bm{E}_{e}\times\bm{H}_{e}\cdot\hat{\bm{n}}=\frac{a_{e}^{2}}{2}\int\limits_{p}da\ \bm{e}_{p}\times\bm{h}_{p}\cdot\hat{\bm{n}}=\frac{\alpha_{i}^{2}\omega_{i}u_{i}}{Q_{e}}. \label{eq:useful4}
\end{equation}


\section{\label{Lagrangian} A Variational Formulation For the Beam-Loaded Cavity}
\subsection{\label{Outline} Method's outline}
For the cavity setup described in the preceding section, we now proceed to write a suitable Lagrangian, $\mathcal{L}$, that will capture the system dynamics \cite{zangwill,Goldstein}. We follow an approach similar to Schwinger's \cite{schwinger} in deriving a variational formulation for a set of time-harmonic fields which must meet Maxwell's equations and the boundary conditions in hand. For time-harmonic fields with an action integral $I{=}\int^{t_{2}}_{t_{1}}\mathcal{L} dt$ and a system Lagrangian $\mathcal{L}{=}\int_{V}dv \tilde{L}(x,y,z,\bm{E},\bm{H})$, where $\tilde{L}$ is the Lagrangian density, we can see that the action's stationary property for a field variation such as $\delta \bm{E}$ or $\delta \bm{H}$ will lead to the stationarity of the Lagrangian; $\delta\mathcal{L}=0$ \cite{schwinger}. For the problem in hand, one would expected $\mathcal{L}$ to generally include terms related to energy density contributions from the field's stored energy, wall losses, beam--field interaction and power flow through the port. Once suitably constructed, $\mathcal{L}$'s stationarity will then automatically settle on the correct solutions that meet the system's equations of motions, which are, for an electromagnetic field, identically equal to Maxwell equations and the imposed boundary conditions \cite{schwinger,zangwill}. 

Since the $E$ and $H$ fields are linked by any of Maxwell's curl equations, we may also construct $\mathcal{L}$ to be solved simultaneously with one of those curl equations, effectively rendering the Lagrangian variation a function of one field variation only (either $\delta\bm{E}$ or $\delta\bm{H}$) \cite{schwinger}. This approach is useful for practical situations, such as the cavity problem under consideration, since the ideal boundary condition at the surface $\partial V$ is a Dirichlet condition \cite{Morse} that may be written as $\hat{\bm{n}}\times\bm{E}=\bm{0}$ or $\hat{\bm{n}}\times\delta\bm{H}=\bm{0}$, forcing a surface integral of the type $\int_{\partial V}da\ \delta\bm{H}\cdot \hat{\bm{n}}\times\bm{E}$ to vanish automatically whenever it appears following the variation of $\mathcal{L}$ \cite{schwinger}. In our cavity, where the walls are highly conductive, it is convenient to work in terms of the magnetic field $\bm{H}$ as the basic variable in $\mathcal{L}$, while the electric field is written in terms of $\bm{H}$ via Maxwell's curl equation, 
\begin{equation}
    \bm{E}=\frac{i}{\omega\epsilon}(\nabla\times\bm{H}-\bm{J}). \label{eq:HAsMainVariable}
\end{equation}

The stationarity of $\mathcal{L}$ in the absence of the port, beam and other losses (undriven) would naturally lead to the wave equation which the eigenmodes ($\bm{e}_{i},\bm{h}_{i}$) satisfy.   Adding the effect of the wall losses will then perturb the fields by effectively allowing them to expand in volume into the skin depth layer of the walls. Under such perturbation, the frequency and $Q$ factor are slightly reduced due to the skin depth effect, while the magnetic field (or surface current) tangent to the walls is approximately unchanged \cite{pozar,slaterbook}, which is equivalent to the variational condition $\hat{\bm{n}}\times\delta\bm{H}{=}\bm{0}$ on $\partial V$, \cite{schwinger}. Indeed, adding the wall losses to the undriven Lagrangian formulation then forcing the Lagrangian to be stationary will be seen to lead to the expected shift in frequency and $Q$ as predicted by classical treatments, e.g.~\cite{slaterbook}. Adding the beam current density and the port will further equip the Lagrangian (driven) to include the necessary nonlinear interaction between the fields and the beam, and allow the calculations to be written in terms of the port's input reflection coefficient. The latter is useful to monitor the shift in frequency and incurred impedance mismatch due to the beam loading, at the stationary point of the Lagrangian. As discussed in Subsection~\ref{geometry} for the adopted single-mode analysis, the field variation $\delta\bm{H}$, and therefore the perturbed complex amplitude $\alpha_{i}$, will hold the information we seek concerning the effect of beam loading on the rf cavity's field, when it is at a stationary point (local minimum) of the Lagrangian in the neighborhood of the eigenmode $(\bm{e}_{i},\bm{h}_{i})$.  In the presented formulation, we have the advantage of working directly in terms of measurable parameters, such as the input power $P_{\text{in}}$, coupling coefficient $\beta_{c}$ and frequency shift $\Delta\omega$.

\subsection{\label{CavityLagrangian} Cavity Lagrangian formulation}

In conjunction with (\ref{eq:HAsMainVariable}), we choose the following Lagrangian construction for the cavity setup under consideration (see Fig~\ref{fig:cavity})
\begin{eqnarray}
\mathcal{L}&=&\frac{1}{2}\int\limits_{V}dV\left( \epsilon E^{2}+\mu H^{2} \right)+\frac{i}{2\omega}\int\limits_{p}da\ \left( \bm{E}_{e}\times\bm{H}_{e} \right)\cdot \hat{\bm{n}} \nonumber \\
&& +\frac{1+i}{2\omega}\xi\int\limits_{\partial V}da\ \bm{H}^{2} +\frac{2i}{\omega}\int\limits_{p}da\ \left( \bm{E}_{e}\times\bm{H}^{+} \right)\cdot \hat{\bm{n}}. \label{eq:L}
\end{eqnarray}

Note that in this formulation of the Lagrangian, the term $\epsilon E^{2}$ appears added to the term $\mu H^{2}$, in contrast to the typical form of the Lagrangian that has the stored electrical energy subtracted from the stored magnetic energy. The apparent change of sign in the current formulation is due to the fact that $E$ and $H$ are time-harmonic phasors and that here $E^{2}$ denotes $\bm{E}\cdot \bm{E}$ (not $\bm{E}\cdot\bm{E}^{*}$, which carries no phase), with a similar meaning for $H^{2}$, as mentioned in Section~\ref{Q_definitions}. This Lagrangian form, which uses the fields themselves rather than the corresponding potential functions, was first adopted by Schwinger \cite{schwinger} to allow one to write a Lagrangian for a dissipative system.  Indeed, a quick consistency test using an arbitrary variation $\delta\bm{H}$ can reveal (see Appendix~\ref{Appx2}) that we are able to automatically recover, from the stationarity of this $\mathcal{L}$, Maxwell's equation for $\bm{\nabla\times\bm{E}}$ as well as the correct boundary conditions on the walls, (\ref{eq:BConWalls}), and on the port's plane, where the tangential components must match those propagating through the guide ($\left. \bm{E}\right|_{p}=\bm{E}_{e}$, $\left. \bm{H}\right|_{p}=\bm{H}_{e}+2\bm{H}^{+}$). 

Another consistency test (see Appendix~\ref{Appx3}) can be done using the undriven $\mathcal{L}$ while writing $\bm{H}$ according to the perturbative approximation given in (\ref{eq:EandH}), namely $\bm{H}\cong \alpha_{i}\bm{h}_{i}$, and taking into account the cavity's wall losses. This immediately shows that the real frequency shift (call it $\Delta\omega_{i}=\omega'_{i}-\omega_{i}$) due to the finite-conductivity of the cavity walls is in agreement with estimates from classical treatments (e.g. \cite{slaterbook}), and is given by
\begin{equation}
   \left(\frac{\omega'_{i}}{\omega_{i}}-\frac{\omega_{i}}{\omega'_{i}}\right)=-\frac{1}{Q_{i}}. \label{eq:FreqShiftInitial}
\end{equation}
This can be simplified for high-$Q_{i}$ cavities as
\begin{eqnarray}
-\frac{1}{Q_{i}}&=&\left(\frac{\omega'_{i}}{\omega_{i}}-\frac{\omega_{i}}{\omega'_{i}}\right)\cong\frac{2\Delta\omega_{i}}{\omega'_{i}}, \\
\Rightarrow  \omega'_{i}&=&\omega_{i}\left(1-\frac{1}{2Q_{i}}\right),   \label{eq:omega_r}
\label{eq:omega_prime}
\end{eqnarray}
which shows, as expected \cite{slaterbook}, that the resonant frequency correction for the eigenmode is small for high-$Q$ cavities.  

It will be convenient to generally introduce a parameter, $\delta_{\omega}$, defined as
\begin{eqnarray}
\delta_{\omega}\equiv\left(\frac{\omega}{\omega_{i}}-\frac{\omega_{i}}{\omega}\right), \label{eq:delta}
\end{eqnarray}
to represent the relative shift in a frequency $\omega$ compared to the resonant frequency $\omega_{i}$. In the above case, for the undriven cavity with finite-conductivity walls, it is easy too see that $\delta_{\omega}$ will be equal to $-1/Q_{i}$.  

The inclusion of the beam and the port terms in (\ref{eq:L}) will further shift the frequency beyond the shift measured due to the finite conductivity of the cavity walls. Therefore, we can choose our frequency reference in practice to be that of the cavity with realistic finite-conductivity walls, $\omega'_{i}$, as given by (\ref{eq:omega_r}). Using primes in the notation, we can extend this to the general definition of $\delta_{\omega}$ as well. In this ``primed" regime, all frequency shifts are measured relative to the undriven cavity with finite-conductivity walls. It is easy to show that $\delta_{\omega}'$ in this regime is given by 
\begin{equation}
\delta'_{\omega}=\delta_{\omega}+\frac{1}{Q_{i}}, \ \ \ \ \delta'_{\omega}\equiv\left( \frac{\omega}{\omega'_{i}}-\frac{\omega'_{i}}{\omega} \right)\cong \frac{2\Delta\omega'}{\omega}. \label{eq:delta_prime}
\end{equation}

We now substitute the perturbative modal representation (\ref{eq:curl_ei})--(\ref{eq:EandH}) and the parameter definitions derived in Section~\ref{Prelim} to obtain the inhomogeneous $\mathcal{L}$ for the driven cavity, including the field--beam interaction, whose stationarity will lead to the correct dynamical law of $\alpha_{i}$. Starting with (\ref{eq:L}), then substituting from (\ref{eq:ui}) and (\ref{eq:useful1})--(\ref{eq:useful4}), and using (\ref{eq:HAsMainVariable}) to work in terms of the main variable $\bm{H}$, one can see (after some manipulation) that

\begin{widetext}
\begin{eqnarray}
\mathcal{L}&=&\frac{1}{2}\int\limits_{V}dV\left( \epsilon E^{2}+\mu H^{2} \right) +\frac{1+i}{2\omega}\xi\int\limits_{\partial V}da\ H^{2} +\frac{i}{2\omega}\int\limits_{p}da\ \left( \bm{E}_{e}\times\bm{H}_{e} \right)\cdot \hat{\bm{n}}+\frac{2i}{\omega}\int\limits_{p}da\ \left( \bm{E}_{e}\times\bm{H}^{+} \right)\cdot \hat{\bm{n}} \nonumber \\
&=&\int\limits_{V}dV\frac{-1}{2\omega^{2}\epsilon}\left(\alpha_{i}\nabla\times\bm{h}_{i}-\bm{J} \right)^{2}+\frac{\alpha_{i}^{2}}{2}\int\limits_{V}dV\mu h_{i}^{2}+ \frac{(1+i)\alpha_{i}^{2}}{2\omega}\frac{\omega_{i}}{Q_{i}}\int\limits_{V}dV \mu h_{i}^{2}+i\frac{\omega_{i}}{\omega}\frac{\alpha_{i}^{2}u_{i}}{Q_{e}}-i\frac{4\alpha_{i}}{\omega}e^{i\phi^{+}}\sqrt{\frac{P_{\textrm{in}}\omega_{i}u_{i}}{Q_{e}}} \nonumber \\
&=&\alpha_{i}^{2}u_{i}\left[ 1-\left( \frac{\omega_{i}}{\omega} \right)^{2} \right]-\frac{1}{2\epsilon\omega^{2}}\int\limits_{V}dVJ^{2}-i\frac{\alpha_{i}\omega_{i}}{\omega^{2}}\int\limits_{V}dV \bm{e}_{i}\cdot\bm{J} + \alpha_{i}^{2}\frac{(1+i)}{\omega}p_{i}+i\frac{\alpha_{i}^{2}}{\omega}p_{i}\beta_{c}-i\frac{4\alpha_{i}}{\omega}e^{i\phi^{+}}\sqrt{P_{\textrm{in}}\beta_{c}p_{i}}, 
\end{eqnarray}
whose stationarity in $\alpha_{i}$ requires
\begin{eqnarray}
\delta\mathcal{L}=0&=&2i\alpha_{i}p_{i}\left[  -Q_{i}\delta_{\omega}-(1+i)-i\beta_{c} \right]- \frac{\omega_{i}}{\omega}\int\limits_{V}dV\bm{e}_{i}\cdot\bm{J}-4e^{i\phi^{+}}\sqrt{P_{\textrm{in}}p_{i}\beta_{c}},
\end{eqnarray}
\end{widetext}
leading us to the dynamical law of $\alpha_{i}$ that we seek
\begin{equation}
    \alpha_{i}=\frac{2e^{i\phi^{+}}\sqrt{P_\textrm{in}\beta_{c}/p_{i}}+\frac{1}{2p_{i}}\frac{\omega_{i}}{\omega}\int_{V}dV \bm{e}_{i}\cdot\bm{J}}{(1+\beta_{c})-i(1+\delta_{\omega}Q_{i})}. \label{eq:alpha}
\end{equation}

The important result given by (\ref{eq:alpha}) can be recast, for a high-$Q_{i}$ cavity, in terms of the primed frequency regime and $Q_{t}$ by making use of (\ref{eq:omega_prime}), (\ref{eq:delta_prime}) and (\ref{eq:QtinTermsOfQi}), to yield
\begin{equation}
    \alpha_{i}\cong\frac{2e^{i\phi^{+}}\sqrt{P_\textrm{in}\beta_{c}/p_{i}}+\frac{1}{2p_{i}}\frac{\omega'_{i}}{\omega}\int_{V}dV \bm{e}_{i}\cdot\bm{J}}{(1+\beta_{c})[1-i\delta'_{\omega}Q_{t}]}. \label{eq:alphainPrimed}
\end{equation}

Let us define the port's steady-state reflection coefficient at the port's plane by
\begin{eqnarray}
\Gamma&=&\frac{E_{\text{ref.}}}{E_{\text{inc.}}}=\left.\frac{E^{-}+E_{e}}{E^{+}}\right|_{p}\cong\frac{a_{e}-a^{+}}{a^{+}} \nonumber\\
&&=\frac{a_{e}}{a^{+}}-1=e^{-i\phi^{+}}\alpha_{i}\sqrt{\frac{p_{i}\beta_{c}}{P_{\text{in}}}}-1. \label{eq:useful5}
\end{eqnarray}

A quick consistency test of the $\alpha_{i}$ law in (\ref{eq:alphainPrimed}) may now be executed by considering an unloaded cavity ($\bm{J}$=0), which would lead to a reflection coefficient given by
\begin{eqnarray}
    \Gamma&=&e^{-i\phi^{+}}\alpha_{i}\sqrt{\frac{p_{i}\beta_{c}}{P_\textrm{in}}}{-}1=\frac{2\sqrt{P_\textrm{in}\beta_{c}/p_{i}}}{(1+\beta_{c})[1-i\delta'_{\omega}Q_{t}]} \sqrt{\frac{p_{i}\beta_{c}}{P_\textrm{in}}}{-}1 \nonumber \\
    &{=}& \frac{(\beta_{c}-1)+i\delta'_{\omega}Q_{t}(1+\beta_{c})}{(1+\beta_{c})-i\delta'_{\omega}Q_{t}(1+\beta_{c})} {=}-\frac{(1-\beta_{c})-i\delta'_{\omega}Q_{i}}{(1+\beta_{c})-i\delta'_{\omega}Q_{i}}. \label{eq:Gamma_unloadedCase}
\end{eqnarray}

Setting (\ref{eq:Gamma_unloadedCase}) equal to zero (critical matching) will, indeed, lead to the expected values of $\beta_{c}=1$ and $\delta'_{\omega}\cong 2\Delta\omega'/\omega'_{i}=-1/Q_{i}$ for the unloaded cavity.

We now proceed to calculate the integral term, $\int_{V}dV\bm{e}_{i}\cdot\bm{J}$, in (\ref{eq:alpha}) and (\ref{eq:alphainPrimed}).  A current density $\bm{\mathbb{J}}$ that represents a travelling charge density is generally related to the charge density $\rho$ and the relative velocity vector, $\bm{\beta}=\bm{v}/c$, by $\bm{\mathbb{J}}=\rho c\bm{\beta}$, where $c$ is the speed of light and $\bm{v}$ is vector velocity. Describing a collection of charged particles as a point-charge sum (also known as a Klimontovich sum \cite{FELbook}),
\begin{equation}
    \rho=\sum\limits_{j}q_{j}\delta(x-x_{j})\delta(y-y_{j})\delta(z-z_{j}),
\end{equation}
where $q_{j}$ is the $j$-th particle charge in Coulomb and $\delta(\cdot)$ denotes Dirac's delta function, leads to writing
\begin{equation}
    \bm{\mathbb{J}}=c\sum\limits_{j} q_{j}\bm{\beta}_{j}\delta(x-x_{j})\delta(y-y_{j})\delta(z-z_{j}).
\end{equation}

For the current harmonic-field treatment, it can be shown through Fourier analysis that the harmonic current component $J$, at the same frequency $\omega$ as the $(\bm{E}, \bm{H})$ fields, is related to the time-domain representation $\bm{\mathbb{J}}$ by
\begin{eqnarray}
    &&\bm{J}=\bm{J}(\omega)=\frac{2}{T}{\int\limits_{0}^{T}} dt\ \bm{\mathbb{J}}\ e^{i\omega t} \nonumber \\
    &&=\frac{2 c}{T}\sum\limits_{j}q_{j}\int\limits_{0}^{T} dt \bm{\beta}_{j}\delta(x{-}x_{j})\delta(y{-}y_{j})\delta(z{-}z_{j})e^{i\omega t}, \label{eq:Jfourier}
\end{eqnarray}
where $T=2\pi/\omega$ is the field's period.

In anticipation of the typical application of this analysis to axisymmetric structures, such as reentrant cavity gaps and beam tubes, it is convenient and customary to work in terms of the axial dimension $z$ as the independent variable (instead of $t$). From the properties of Dirac's delta function for the $j$-th particle's axial displacement, and since $|v_{z,j}|=v_{z,j}$ here, we notice that $\delta(z-z_{j})=\delta(v_{z,j}\Delta t_{j})=\frac{1}{v_{z,j}}\delta(\Delta t_{j})=\frac{1}{c\beta_{z,j}}\delta(t-t_{j})$. Equation (\ref{eq:Jfourier}) can now be rewritten as
\begin{equation}
    \bm{J}=\frac{2}{T}\sum\limits_{j}q_{j}\int\limits_{0}^{T} dt \frac{\bm{\beta}_{j}}{\beta_{z,j}}\delta(x{-}x_{j})\delta(y{-}y_{j})\delta(t{-}t_{j})e^{i\omega t}, \label{eq:Jfourier_z}
\end{equation}
where $\beta_{z}$ is the $z$-component of the vector $\bm{\beta}$.

We can now utilize (\ref{eq:Jfourier_z}) in the integral $\int_{V}dV\bm{e}_{i}\cdot\bm{J}$, and drop the mode subscript $i$ in $\bm{e}_{i}$ for simplicity, to write
\begin{eqnarray}
    &&\int\limits_{V}dV \bm{e}\cdot\bm{J}= \nonumber\\
    &&\frac{2}{T}\sum\limits_{j}q_{j}\int\limits_{0}^{T}\int\limits_{V} dtdV\  \bm{e}\cdot\frac{\bm{\beta}_{j}}{\beta_{z,j}}\delta(x{-}x_{j})\delta(y{-}y_{j})\delta(t{-}t_{j})e^{i\omega t} \nonumber \\
    &&=\frac{2}{T}\sum\limits_{j} q_{j} \int\limits_{z_{i}}^{z_{f}}dz\ \bm{e}_{j}\cdot\frac{\bm{\beta}_{j}}{\beta_{z,j}} e^{i\omega t_{j}},  \label{eq:IntegralReady}
\end{eqnarray}
where $z_{i}, z_{f}$ are the particle's initial and final axial positions in the interaction zone. The electric field $\bm{e}_{j}=\bm{e}_{j}(x_{j},y_{{j}},t_{j})$ in (\ref{eq:IntegralReady}) is now evaluated at the transverse position $(x_{j},y_{j})$ and time $t_{j}$ which the $j$-th particle will have when it reaches the distance $z$ down the axis. As the $j$-th particle follows its own trajectory (which will be later described by the equations of motion), the time $t_{j}$ will play a key role in describing the phase of the rf field that will be encountered by the particle upon its entry to the zone of interaction with the field (the cavity's gap). This phase can be described in terms of the time variable $t_{j}$ or, as will be seen in the next section, in terms of an rf field phase variable, $\phi_{j}$, seen by the particle.

The equation of motion for the $j$-th particle due to the experienced electric Lorentz force gives 
\begin{equation}
    \frac{d\gamma_{j}}{dt}=\frac{e_{c}}{m c^{2}}\textrm{Re}\left( \bm{E}_{j} e^{-i\omega t_{j}}\right), \label{eq:dg/dt_initially}
\end{equation}
where $e_{c}$ is the electron charge, $m$ is the electron rest mass and $\gamma$ is Lorentz's relativistic factor. The quantity $m c^{2}/e_{c}$ is denoted $V_{0}$, which is approximately $0.511$ MV. Let us now introduce a convenient complex function, $\gamma^{c}_{j}\equiv\gamma_{j}+i\breve{\gamma}_{j}$, whose real part is the usual Lorentz relativistic factor and imaginary part is yet undetermined. Using (\ref{eq:dg/dt_initially}) and the fact that $\textrm{Re}\left ( \bm{E}e^{-i\omega t} \right)=\textrm{Re}\left ( \bm{E}^{*}e^{+i\omega t} \right)$, we may now write an extended version of (\ref{eq:dg/dt_initially}) as
\begin{equation}
    \frac{d\gamma^{c}_{j}}{dt}=\frac{c}{V_{0}}\bm{E}^{*}_{j}\cdot\bm{\beta}_{j} e^{i\omega t_{j}}=\frac{c}{V_{0}}\alpha^{*}_{i}\bm{e}^{*}_{j}\cdot\bm{\beta}_{j} e^{i\omega t_{j}}, \label{eq:dg_dt_complex}
\end{equation}
which implies that
\begin{equation}
\frac{d\gamma^{c}_{j}}{dz}=-\frac{\alpha_{i}^{*}}{V_{0}}\bm{e}_{j}\cdot\frac{\bm{\beta}_{j}}{\beta_{z,j}} e^{i\omega t_{j}}. \label{eq:dg_dz_complex} 
\end{equation}

Using (\ref{eq:dg_dz_complex}) in (\ref{eq:IntegralReady}) gives the useful relation
\begin{eqnarray}
    \int\limits_{V}dV \bm{e}\cdot\bm{J}&=&-\frac{2V_{0}}{T\alpha_{i}}e^{2i\phi_{\alpha_{i}}}\sum\limits_{j}q_{j}\int\limits_{z_{i}}^{z_{f}}\frac{d\gamma^{c}_{j}}{dz}dz \nonumber \\
    &=&-\frac{2V_{0}}{T\alpha_{i}}e^{2i\phi_{\alpha_{i}}}\sum\limits_{j} q_{j} \Delta\gamma^{c}_{j},
\end{eqnarray}
which is substituted back into $\alpha_{i}$'s law, (\ref{eq:alphainPrimed}), to yield
\begin{equation}
    \alpha_{i}\cong\frac{2e^{i\phi^{+}}\sqrt{P_\textrm{in}\beta_{c}/p_{i}}-\frac{1}{p_{i}}\frac{\omega'_{i}}{\omega}\frac{V_{0}}{T\alpha_{i}^{*}}\sum_{j} q_{j} \Delta\gamma^{c}_{j}}{(1+\beta_{c})[1-i\delta'_{\omega}Q_{t})}. \label{eq:alphainPrimedinGamma}
\end{equation}

 Equation (\ref{eq:alphainPrimedinGamma}) constitutes the sought dynamical law governing $\alpha_{i}$, now written in terms  of the particles' $\gamma^{c}_{j}$ functions, alongside the other system parameters, and in a form that is ready for computation. In words, (\ref{eq:alphainPrimedinGamma}) gives the steady-state's complex amplitude $\alpha_{i}$ corresponding to a given cavity--beam setup that has a frequency shift parameter $\delta_{\omega}$ and coupling coefficient value $\beta_{c}$. Up to this point the discussion has been general and the developed theory is applicable to an arbitrary cavity. The calculation of the sum involving $\Delta\gamma$ in (\ref{eq:alphainPrimedinGamma}) may be further simplified for computation by considering a specific configuration, as will be discussed in the next section.  

\begin{figure}
\includegraphics[width=8.5cm]{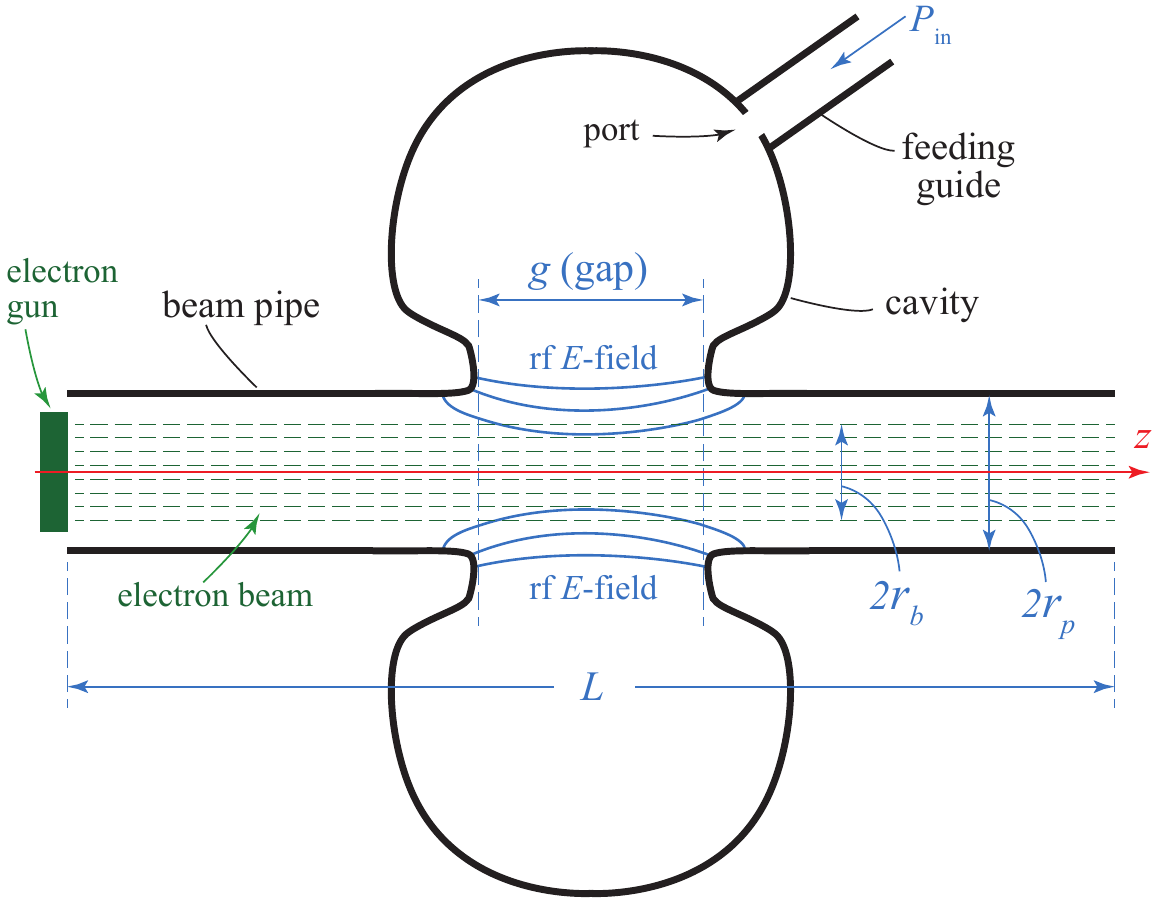}
\caption{\label{fig:inputcavity} A simplified drawing of the geometry of a klystron input cavity. The typical reentrant cavity shape forms a gap (ungridded) where the electric field interacts with the beam. The electron gun is represented as a conceptual block.}
\end{figure}


\section{\label{InputCavity} Application of the Theory to A Klystron Input Cavity}
\subsection{\label{XandY} Cavity parameters $\bar{X},\bar{Y}$ for the characterization of beam-loading effects}
We apply the theory developed in the preceding sections to an important class of systems that is exemplified by a klystron input cavity. In such a system, the cavity will typically have an azimuthally-symmetric reentrant shape, with axis $z$ that aligns with the beam pipe, as illustrated in  Fig~\ref{fig:inputcavity}. The gap approximately defines the zone where the cavity's rf field and the beam will interact. Away from the interaction zone, the beam pipe's elongations and small apertures at the start and end sections of the pipe (typically leading to other drift tubes or adjacent cavities) will not have any significant coupling effect on the cavity's fields, since it is assumed that any evanescent rf fields would have died off away from the gap. As such, the cavity--beam interaction may still be modelled by the single-port cavity theory developed in Section~\ref{Lagrangian}. Given a dc beam voltage $V_{b}$, the electrons exiting the electron gun will have an initial $\gamma_{0}$ value of $\gamma_{0}=1+V_{b}/V_{0}$, \cite{Reiser}.  Moreover, since we have the freedom to normalize $u_{i}$, let us normalize it such that the post-normalization value of $p_{i}$ is equal to the beam's dc power; namely, $p_{b}=V_{b}I_{b}=p_{i}=\omega_{i}u_{i}/Q_{i}$, where $I_{b}$ is the beam current. Under such conditions, $|\alpha_{i}|$ can be treated as a small parameter from the beam's perspective, and the field interaction with the travelling particle is viewed as a perturbation problem in which the field's nonzero magnitude $|\alpha_{i}|$ tends to perturb the trajectory of the particle (compared to its unperturbed trajectory when the rf fields are absent). This will be convenient in the following perturbation calculations.

Substituting the normalized values of $u_{i}$ and $(\gamma_{0}-1)=V_{b}/V_{0}$ into (\ref{eq:alphainPrimedinGamma}) gives
\begin{equation}
    \alpha_{i}\cong\frac{2e^{i\phi^{+}}\sqrt{P_\textrm{in}\beta_{c}/p_{b}}-\frac{1}{I_{b}T\alpha_{i}^{*}(\gamma_{0}-1)}\frac{\omega'_{i}}{\omega}\sum_{j} q_{j} \Delta\gamma^{c}_{j}}{(1+\beta_{c})[1-i\delta'_{\omega}Q_{t})}. \label{eq:alpha_temp1}
\end{equation}

Replacing the cumbersome Klimontovich sum of point-like particles by an approximate macro-particle distribution (over sufficiently small bins in phase space) will simplify the calculation.  To this end, the charges $q_{j}$ can be represented as $q_{j}=q_{T}W_{j}$, where $q_{T}=I_{b}T$ is the total charge crossing in the time period $T$ and $W_{j}=W_{j}\left( r_{j}(z),t_{j}(z);z \right)$ is a spatial weighing function that determines the equivalent charge representation for the $j$-th macro-particle. Note that the index $j$ is now used to refer to the macro-particles and not the original point-like particles. Equation (\ref{eq:alpha_temp1}) can now be written in terms of $W_{j}$ as
\begin{equation}
    \alpha_{i}\cong\frac{2e^{i\phi^{+}}\sqrt{P_\textrm{in}\beta_{c}/p_{b}}-\frac{1}{\alpha_{i}^{*}(\gamma_{0}-1)}\frac{\omega'_{i}}{\omega}\sum_{j} W_{j} \Delta\gamma^{c}_{j}}{(1+\beta_{c})[1-i\delta'_{\omega}Q_{t})}. \label{eq:alpha_temp2}
\end{equation}

For the azimuthally-symmetric case in hand, the natural division for the macro-particles will be radially (over $r$) and longitudinally (over $t$ or temporal phase). For $n$ radial bins of equal charge, which we shall call ``beamlets", and assuming that the initial charge density is uniform in the beam pipe, every beamlet will be located within a radial slice from radius $a_{j}$ to radius $a_{j+1}$, with $j=0, 1, \cdots, n$,  $a_{0}=0$ and $a_{n}=r_{b}$, where $r_{b}$ is the beam radius. To get the same charge per beamlet, one must choose the beamlet radii such that they give the same volume; $a_{j+1}^{2}-a_{j}^{2}=r_{b}^{2}/n$. This gives the solution  $a_{j}=r_{b}\sqrt{j/n}$. The midpoint radius of each beamlet is chosen to be the center-of-mass of charge density distribution between $a_{j}$ and $a_{j+1}$, which gives $a_{j+1}^{2}-r_{j}^{2}=r_{j}^{2}-a_{j}^{2}$. This leads to the beamlet radii 
\begin{equation}
    r_{j}=r_{b}\sqrt{\frac{2j+1}{2n}}, \ \ \ (j=0, 1, \cdots, n-1). \label{eq:beamlet_radius}
\end{equation}

Since the charge is distributed over all times (phases) within a period $T=2\pi/\omega$, a longitudinal division into $m$ time bins ($t_{k}=k \Delta t=kT/m$) or $m$ phase bins ($\phi_{k}=k2\pi/m$) is considered, where $k$ is a running index, $k=0, 1, \cdots, m$.  The macro-charge summing term $\sum_{j} W_{j} \Delta\gamma^{c}_{j}$ in (\ref{eq:alpha_temp2}) may now be written as an averaging process, denoted $\langle \Delta\gamma^{c} \rangle_{j,k}$, over the beamlets and phases, with the understanding that the index $j$ now runs over the beamlet radii of the macro-particles, while the index $k$ runs over their phase bins;
\begin{equation}
    \sum_{j} W_{j} \Delta\gamma^{c}_{j}{\equiv}\frac{1}{mn}\sum\limits_{k=0}^{m-1}\sum\limits_{j=0}^{n-1}\Delta\gamma^{c}(r_{j},\phi_{k})\equiv\langle \Delta\gamma^{c} \rangle_{j,k}. \label{eq:AverageSum}
\end{equation}

Under this representation, (\ref{eq:alpha_temp2}) is written as
\begin{equation}
    \alpha_{i}\cong\frac{2e^{i\phi^{+}}\sqrt{P_\textrm{in}\beta_{c}/p_{b}}-\frac{1}{\alpha_{i}^{*}(\gamma_{0}-1)}\frac{\omega'_{i}}{\omega}\langle \Delta\gamma^{c} \rangle_{j,k}}{(1+\beta_{c})[1-i\delta'_{\omega}Q_{t})}, \label{eq:alpha_temp3}
\end{equation}
where the term $\langle \Delta\gamma^{c} \rangle_{j,k}$ may now be computed from Hamilton's equations of motion.

Looking back at the expression for $d\gamma^{c}/dz$ for the macro-particle's $j$-th beamlet and $k$-th phase in light of  (\ref{eq:dg_dz_complex}), we see that it is given explicitly by
\begin{equation}
 \frac{d\gamma^{c}_{j,k}}{dz}=-\frac{|\alpha_{i}|}{V_{0}}\bm{e}(r_{j},t_{k})\cdot\frac{\bm{\beta}(r_{j},t_{k})}{\beta_{z}(r_{j},t_{k})} e^{i(\omega t +\phi_{k})} e^{-i\phi_{\alpha_{i}}}, \label{eq:dg_dz_complex_macro} 
\end{equation}
where $t_{k}\equiv t+\phi_{k}/\omega$. Recall that $u_{i}$ was normalized in a way that makes $|\alpha_{i}|$ a small perturbation parameter, describing how the rf field would disturb the particle trajectory as it travels through the gap. Using perturbation theory to characterize the behavior of $\gamma^{c}$, we expand the phase variables $r(z)$, $t(z)$ and the functions $\gamma_{c}$, $\beta$, which appear in (\ref{eq:dg_dz_complex_macro}), as power series in the small parameter $|\alpha_{i}|$. If we momentarily drop the subscripts $i, j, k$, for clarity during this calculation, we can write the expansions up to second order in $|\alpha_{i}|$ as
\begin{eqnarray}
    \gamma^{c}&=&\gamma^{c}_{0}+|\alpha|\gamma^{c}_{1}+|\alpha|^{2}\gamma^{c}_{2}+\mathcal{O}(|\alpha|^{3}), \\
    \beta&=&\beta_{0}+|\alpha|\beta_{1}+|\alpha|^{2}\beta_{2}+\mathcal{O}(|\alpha|^{3}), \\
    r&=&r_{0}+|\alpha|r_{1}+|\alpha|^{2}r_{2}+\mathcal{O}(|\alpha|^{3}), \\
    t&=&t_{0}+|\alpha|t_{1}+|\alpha|^{2}t_{2}+\mathcal{O}(|\alpha|^{3}),
\end{eqnarray}
where the first terms represent the unperturbed problem when $\alpha_{i}=0$ (no rf fields). Substituting the expansions into (\ref{eq:dg_dz_complex_macro}) and equating terms of similar powers on both sides, while using the initial codition $\gamma^{c}_{0}=\gamma_{0}$ and the notation $\Delta\gamma^{c}\equiv\gamma^{c}-\gamma_{0}$, it can be shown that 
\begin{eqnarray}
    \gamma^{c}_{0}&=&\gamma_{0}, \\
    \frac{d\gamma^{c}_{1}}{dz}&=&-\frac{1}{V_{0}}\bm{e}(r_{0},t_{0})\cdot\frac{\bm{\beta}(r_{0},t_{0})}{\beta_{z}(r_{0},t_{0})} e^{i(\omega t +\phi_{k})} e^{-i\phi_{\alpha}}, \\
    \frac{d\Delta\gamma^{c}_{1}}{dz}&=&-\frac{1}{V_{0}}\bm{e}(r_{0},t_{0})\cdot\frac{\bm{\beta}(r_{0},t_{0})}{\beta_{z}(r_{0},t_{0})} e^{i(\omega t +\phi_{k})} e^{-i\phi_{\alpha}} \label{eq:DeltaGamma1},\\
    \Delta\gamma^{c}&=&|\alpha|\Delta\gamma^{c}_{1}+|\alpha|^{2}\Delta\gamma^{c}_{2}+\mathcal{O}(|\alpha|^{3}),
\end{eqnarray}
where $r_{0}, t_{0}, \beta_{0}$ represent the trajectories of the unperturbed problem. Consequently, the average of (\ref{eq:DeltaGamma1}) over the phases $\phi_{k}$ of the field will vanish, implying that only the second order and higher terms will contribute to the average of $\Delta\gamma^{c}$. Restoring the subscript notation $i,j,k$, this observation may be now written as
\begin{equation}
    \langle \Delta\gamma^{c} \rangle_{k,j}=|\alpha_{i}|^{2}\langle \Delta\gamma^{c}_{2} \rangle +\mathcal{O}(|\alpha_{i}|^{3}), \label{eq:SecondOrderOnlySurvives}
\end{equation}
giving the dynamic law of $\alpha_{i}$ in (\ref{eq:alpha_temp3}), up to second order in $|\alpha_{i}|^{2}$, as
\begin{eqnarray}
    \alpha_{i}&\cong&\frac{2e^{i\phi^{+}}\sqrt{P_\textrm{in}\beta_{c}/p_{b}}-\frac{1}{\alpha_{i}^{*}(\gamma_{0}-1)}\frac{\omega'_{i}}{\omega}|\alpha_{i}|^{2}\langle \Delta\gamma^{c}_{2} \rangle}{(1+\beta_{c})[1-i\delta'_{\omega}Q_{t})} \nonumber\\
    &=&\frac{2e^{i\phi^{+}}\sqrt{P_\textrm{in}\beta_{c}/p_{b}}-\frac{\alpha_{i}}{(\gamma_{0}-1)}\frac{\omega'_{i}}{\omega}\langle \Delta\gamma^{c}_{2} \rangle}{(1+\beta_{c})[1-i\delta'_{\omega}Q_{t})}. \label{eq:solve_for_alpha}
\end{eqnarray}

Solving (\ref{eq:solve_for_alpha}) for $\alpha_{i}$ and writing $\langle \Delta\gamma^{c}_{2} \rangle/(\gamma_{0}-1)$ explicitly as a complex quantity,
\begin{equation}
\frac{\langle \Delta\gamma^{c}_{2} \rangle}{\gamma_{0}-1}=\bar{X}-i\bar{Y}, \label{eq:XandY}
\end{equation}
gives us the final result
\begin{equation}
    \alpha_{i}=\frac{2 e^{i\phi^{+}}\sqrt{P_\textrm{in} \beta_{c}/p_{b}}}{(1+\beta_{c})(1-\delta'_{\omega}Q_{t})+\frac{\omega_{i}}{\omega}(\bar{X}-i\bar{Y})}.
    \label{eq:final_alpha}
\end{equation}

The significance of the result in (\ref{eq:final_alpha}) stems from the fact that the two paramaters $\bar{X}$ and $\bar{Y}$ are independent from $|\alpha_{i}|$ and only dependent on the rf field distribution (cavity shape) and the beam's initial conditions ($\gamma_{0}$). The parameters $\bar{X}$ and $\bar{Y}$ can thus concisely characterize any cavity setup and how it would detune under beam-loading. 

Equation (\ref{eq:final_alpha}) gives the value of $\alpha_{i}$ corresponding to a given cavity--beam system that is running with a frequency shift parameter $\delta_{\omega}$ and a coupling coefficient value $\beta_{c}$. We can now use this result to suggest a way to pretune (pre-compensate) the cavity in a preemptive manner before loading it with the beam, so that it is tuned back to the nominal ($\omega'_{i}$) resonant frequency and critical coupling when the beam is turned on.  Specifically, the reflection coefficient $\Gamma$ corresponding to the complex amplitude given in (\ref{eq:final_alpha}) can be calculated by substituting (\ref{eq:final_alpha}) into (\ref{eq:useful5}), which yields (after some manipulation)
\begin{equation}
    \Gamma=-\frac{\left( 1-\beta_{c}+\frac{\omega'_{i}}{\omega}\bar{X} \right) -i \left( \delta'_{\omega}Q_{i}+\frac{\omega'_{i}}{\omega}\bar{Y}  \right)}{\left( 1+\beta_{c}+\frac{\omega'_{i}}{\omega}\bar{X} \right) -i \left( \delta'_{\omega}Q_{i}+\frac{\omega'_{i}}{\omega}\bar{Y}  \right)}. \label{eq:final_Gamma}
\end{equation}

Requiring $\Gamma$ to vanish will immediately lead to $\delta'_{\omega}$ and $\beta_{c}$ values given by
\begin{eqnarray}
    \beta_{c}&=&1+\frac{\omega'_{i}}{\omega}\bar{X}, \label{eq:penultimate1}\\
    \delta'_{\omega}Q_{i}&=&-\frac{\omega'_{i}}{\omega}\bar{Y}. \label{eq:penultimate2}
\end{eqnarray}

Using (\ref{eq:delta_prime}) to deduce that $\omega'_{i}/\omega=(1-\delta'_{\omega}/2)$ and substituting in (\ref{eq:penultimate1}) and (\ref{eq:penultimate2}) finally gives, up to second order in $1/Q_{i}$,
\begin{eqnarray}
    \delta'_{\omega}&\cong& -\frac{\bar{Y}}{Q_{i}}-\frac{\bar{Y}^{2}}{2Q_{i}^{2}}, \\
    \beta_{c}&\cong&1+\left( 1+\frac{\bar{Y}}{2Q_{i}}+\frac{\bar{Y}^{2}}{4Q_{i}^{2}} \right)\bar{X}.
\end{eqnarray}

To the lowest order (assuming high-$Q$ cavities) and using practical terminology by calling $\omega'_{i}\equiv\omega_\textrm{cold}$ and $\omega\equiv\omega_\textrm{hot}$, the final result takes the simple form
\begin{eqnarray}
    \delta'_{\omega}\cong-\frac{\bar{Y}}{Q_{i}} \ \ &\Rightarrow& \ \ \omega_\textrm{hot}\cong \omega_\textrm{cold}\left(1-\frac{\bar{Y}}{2Q_{i}}\right), \label{eq:FINAL_f_shift} \\
    \beta_{c}&=&1+\bar{X}. \label{eq:FINAL_beta}
\end{eqnarray}

Equations (\ref{eq:FINAL_f_shift}) and (\ref{eq:FINAL_beta}) are useful for prescribing the frequency shift and coupling coefficient values that would give optimal efficiency (no reflection) in practice, for a given pair of cavity's characterizing parameters, $\bar{X}$ and $\bar{Y}$. A pretuning procedure that maximizes gain based on these results will be discussed in Subsection~\ref{Pretuning}.

Calculating $\bar{X}$ and $\bar{Y}$ for a given system is done using (\ref{eq:XandY}) and  (\ref{eq:SecondOrderOnlySurvives}), which imply that
\begin{equation}
\bar{X}{=}\textrm{Re}\left[\frac{\langle \Delta\gamma^{c} \rangle}{|\alpha_{i}|^{2}(\gamma_{0}-1)}\right], \ \bar{Y}{=}-\textrm{Im}\left[\frac{\langle \Delta\gamma^{c} \rangle}{|\alpha_{i}|^{2}(\gamma_{0}-1)}\right], \label{eq:XandY_FINAL}
\end{equation}
where $\langle \Delta\gamma^{c} \rangle$ is computed, for a given system, by solving the equations of motion. This will be demonstrated numerically in Subsection~\ref{Hamiltonian} and Section~\ref{Examples}. This highlights that the physical processes underlying the detuning of beam-loaded cavities are essentially dynamic and nonlinear. The current formulation also enables us to calculate $\langle \Delta\gamma^{c} \rangle$ under certain conditions without having to solve the equations of motion explicitly, by utilizing canonical transformations and infinitesimal generating functions (e.g.~Lie transformations) which can calculate the value of $\langle \Delta\gamma^{c} \rangle$ analytically (e.g~\cite{Percival,Litchenberg,Lathem,CARY}). As such, the variational formulation allows us to potentially leverage the powerful tools of Hamiltonian dynamics to efficiently solve beam-loading problems. This will be the subject of a subsequent article by the authors.

We note that the detuning parameters $\bar{X}, \bar{Y}$ introduced in this theory, although computed differently, can be modelled as the conventional normalized beam-loading conductance $G_{B}/G$ and susceptance $B_{B}/G$, respectively. Indeed, in light of equations (\ref{eq:Qhot}), (\ref{eq:DeltaF}), (\ref{eq:FINAL_f_shift}) and (\ref{eq:FINAL_beta}), and given that under beam-loaded critical coupling we must have $Q_{e}=Q_{i,\textrm{hot}}=Q_{i,\textrm{cold}}/(1+G_{B}/G)$, a simple algebraic manipulation can show that
\begin{eqnarray}
    \bar{Y}&=&B_{B}/G, \label{eq:link1}\\
    \bar{X}&=&G_{B}/G, \label{eq:link2}\\
    \frac{\Delta Q}{Q_{i,\textrm{cold}}}&=&\frac{-\bar{X}}{1+\bar{X}}.
\end{eqnarray}

One can also calculate the equivalent beam quality factor $Q_\textrm{beam}$, defined  by $1/Q_{i,\textrm{hot}}=1/Q_{i,\textrm{cold}}+1/Q_{\textrm{beam}}$, \cite{Antonsen2002}. Then it is easy to deduce that 
\begin{equation}
    Q_\textrm{beam}=\frac{Q_\textrm{cold}}{\bar{X}}.
\end{equation}

Equations (\ref{eq:link1}) and (\ref{eq:link2}) show that the parameters $\bar{X}, \bar{Y}$ are directly equivalent to a circuit model of a normalized admittance that shunts the cavity to represent the beam-loading effect, as would be expected. The way these parameters are calculated in the present theory [namely, (\ref{eq:final_alpha}) and (\ref{eq:XandY_FINAL})], however, is different from how they are calculated in previous theories, e.g.~\cite{Branch}.

\subsection{\label{Pretuning} Maximizing gain under beam loading via cavity pretuning and shape optimization}

For a given beam-loaded cavity (described by $\bar{X}$ and $\bar{Y}$), the law in (\ref{eq:final_alpha}) is a statement about the complex amplitude $\alpha_{i}$ that the driven cavity field will settle upon in the steady state whilst operating at a given shifted frequency (described by $\delta'_{\omega}$) and a given coupling coefficient (described by $\beta_{c}$). If the cavity is unloaded, then $\bar{X}=\bar{Y}=0$ and (\ref{eq:final_Gamma}) would imply that optimal operation happens when $\beta_{c}=1$ and $\delta'_{\omega}=0 \Rightarrow \omega=\omega'_{i}$ (same frequency as the cavity's original resonant frequency with wall losses included, as expected). If the cavity is loaded, then generally $\bar{X}, \bar{Y}$ are nonzero and (\ref{eq:final_Gamma}) will imply that optimal operation happens at a lower frequency, $\delta'_{\omega}\cong -\bar{Y}/Q_{i} \Rightarrow \omega\cong \omega'_{i}[1-\bar{Y}/(2Q_{i})]$, and higher coupling value, $\beta_{c}=1+\bar{X}$, as given by (\ref{eq:FINAL_f_shift}) and (\ref{eq:FINAL_beta}).

Consequently, if we pretune the cavity to a frequency that is upshifted to $\omega'_{i}[1+\bar{Y}/(2Q_{i})]$ and to an overcoupled port with $\beta_{c}=1+\bar{X}$, we effectively allow beam-tuning to happen in a predetermined way, shifting the frequency down to its original design value $\omega'_{i}$ while guaranteeing that it will be simultaneously matched for maximum power gain. Assuming that the frequency detuning is relatively small, the up or down shifting will result in the same dynamical behavior around the stationary point (sweet spot) of $\alpha_{i}$, and hence one pretuning step may be sufficient (in contrast to iterative tuning steps). These theoretical predictions confirm the empirical observations known in the practice of cavity detuning under beam-loading conditions. In addition to predicting the required values for frequency upshifting and port overcoupling for a given cavity, the present theory can further inform the designer by having the following corollaries:
\begin{enumerate}
    \item[(1)] \emph{A high-$Q$ beam-loaded cavity will have maximum gain in the average Lorentz boost $\langle \Delta \gamma \rangle$ when its pretuning is done to meet the detuning conditions $\delta'_{\omega}=-\bar{Y}/Q_{i}$ and $\beta_{c}=1+\bar{X}$.}
    
    Indeed, this can be proved by noticing from (\ref{eq:SecondOrderOnlySurvives}) and (\ref{eq:XandY_FINAL}) that for high-$Q$ cavities we can simply write \begin{equation}
    \langle \Delta\gamma \rangle \cong \frac{4 P_\textrm{in}\beta_{c}(\gamma_{0}-1)\bar{X}}{p_{b}\left[  (1+\beta_{c}+\bar{X})^{2}+(\delta'_{\omega}Q_{i}+\bar{Y})^{2} \right]} \label{eq:DeltaGammaAverage}
\end{equation}
For any given set of $\bar{X}, \bar{Y}$ parameters, considering $\langle \Delta\gamma \rangle$ as a function of $\delta'_{\omega}$ and $\beta_{c}$ and requiring that both $\partial \langle \Delta\gamma\rangle/\partial \delta'_{\omega}$ and $\partial \langle \Delta\gamma\rangle/\partial \beta_{c}$ vanish, confirms that $(\delta'_{\omega},\beta_{c})=(-\bar{Y}/Q_{i},1+\bar{X})$ represents a critical point of the function $\langle \Delta\gamma \rangle$. Calculating the terms of the Hessian matrix for second-order partial derivatives of $\langle \Delta\gamma \rangle$ in $\delta'_{\omega}$ and $\beta_{c}$, as well as its determinant, $D$, at that critical point for $(\delta'_{\omega},\beta_{c})$ shows that
\begin{eqnarray}
\frac{\partial^{2} \langle \Delta\gamma \rangle }{\partial \delta'^{2}_{\omega}}&<&0, 
\ \ \ \frac{\partial^{2} \langle \Delta\gamma \rangle }{\partial \beta{c}^{2}} <0,
\ \ \ \frac{\partial^{2} \langle \Delta\gamma \rangle }{\partial \delta'_{\omega}\partial \beta{c}} =0, \\
\Rightarrow D&=&\frac{\partial^{2} \langle \Delta\gamma \rangle }{\partial \delta'^{2}_{\omega}} \frac{\partial^{2} \langle \Delta\gamma \rangle }{\partial \beta_{c}^{2}}- \frac{\partial^{2} \langle \Delta\gamma \rangle }{\partial \delta'_{\omega} \partial\beta_{c}}>0,
\end{eqnarray}
which confirms that $\langle \Delta\gamma \rangle$ has maximum gain at the critical point. Substituting these values into (\ref{eq:DeltaGammaAverage}) gives the maximum $\langle \Delta\gamma \rangle$ that can be offered by pretuning, as
\begin{equation}
    \langle \Delta\gamma \rangle_\textrm{max}=\frac{P_\textrm{in}(\gamma_{0}-1)}{p_{b}}\frac{\bar{X}}{1+\bar{X}}. \label{eq:MaxDeltaGammaAve}
\end{equation}
\begin{figure}
\includegraphics[width=7.5cm]{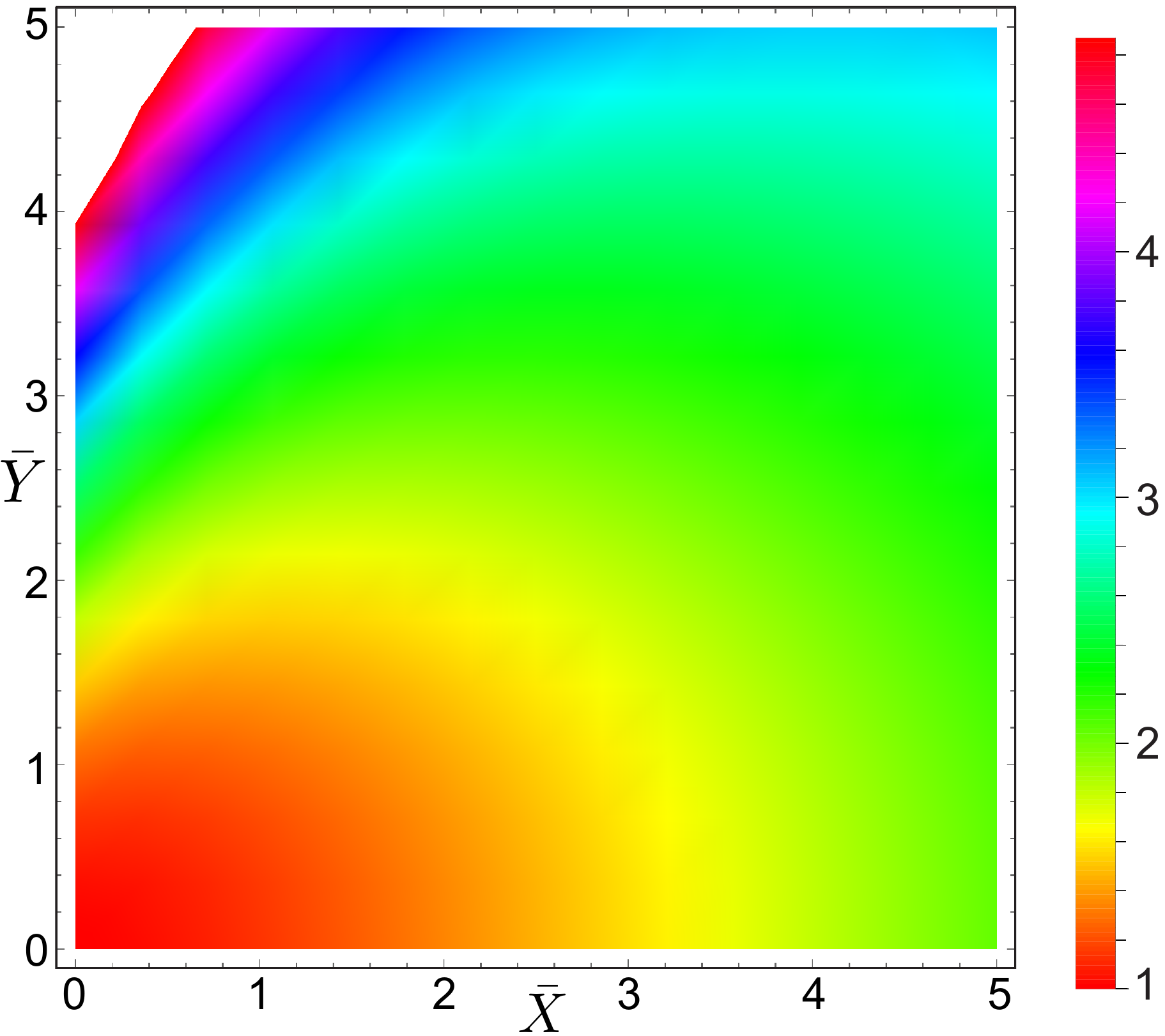}
\caption{\label{fig:fig_3} The maximum detuning gain, $G_\textrm{max}$, as a function of beam-loaded cavity parameters $\bar{X}$ and $\bar{Y}$.}
\end{figure}
If we do not take detuning into account, the value taken by $\langle \Delta\gamma \rangle$ is obtained by substituting $\delta'_{\omega}=0$ and $\beta_{c}=1$ in (\ref{eq:DeltaGammaAverage}). When compared to $\langle \Delta\gamma \rangle_\textrm{max}$, this will give a reduction ratio $R$ as
\begin{equation}
    \frac{1}{\mathcal{G}_\textrm{max}}=R=\frac{\langle \Delta\gamma \rangle_\text{none}}{\langle \Delta\gamma \rangle_\textrm{max}}=\frac{4(1+\bar{X})}{(2+\bar{X})^{2}+\bar{Y}^{2}}, \label{eq:proof}
\end{equation}
where $\mathcal{G}_\textrm{max}$ is introduced to denote the maximum additional gain that our pretuning can achieve in a given system with $\bar{X}, \bar{Y}$ parameters. Requiring $R$ to be $\geq 1$ leads to $0\geq \bar{X}^{2}+\bar{Y}^{2}$, which is impossible to satisfy since $\bar{X}, \bar{Y}$ are real parameters.  Therefore, we always have $R<1$ for any choice of cavity (any $\bar{X}, \bar{Y}$ parameters), as expected in practice.  
    \item[(2)] \emph{For a given system (known $\bar{X}$ and $\bar{Y}$), it follows that the maximum gain obtained in $\langle \Delta\gamma \rangle$ by using the correct pretuning in frequency and couling coefficient is given by
\begin{equation}
    \mathcal{G}_\textrm{max}=\frac{(2+\bar{X})^{2}+\bar{Y}^{2}}{4(1+\bar{X})}. \label{eq:Gain}
\end{equation}} 
Maximizing $\mathcal{G}_\textrm{max}$ can therefore be considered as a design goal during cavity shape optimization, where $\bar{X}$ and $\bar{Y}$ are adjusted as design parameters. Fig~\ref{fig:fig_3} shows a plot for $\mathcal{G}_\textrm{max}$ versus $\bar{X}$ and $\bar{Y}$, highlighting the nonlinear behavior. In Section~\ref{Examples} four numerical examples are presented, demonstrating how cavities of different shapes and beam-loading conditions can vary considerably in how much they boost $\mathcal{G}_\textrm{max}$ upon detuning. 
    \item[(3)] The gain in (\ref{eq:Gain}) has a critical point at $\bar{X}=-1+\sqrt{1+\bar{Y}^{2}}$ that should be avoided if encountered during cavity shape design, as it represents a minimal value for $\mathcal{G}_\textrm{max}$. 
    \item[(4)] \emph{If we only pretune the system in the coupling coefficient we will achieve a suboptimal gain given by
    \begin{equation}
    \mathcal{G}_{c}=\frac{1}{2}+\frac{1}{2}\sqrt{1+\left[ \bar{Y}/(1+\bar{X}) \right]^{2}}. \label{eq:SuboptimalG}
\end{equation}} This can be readily derived by putting $\delta'_{\omega}=0$ and finding the maximum of  (\ref{eq:DeltaGammaAverage}) in terms of $\beta_{c}$.
\end{enumerate}

\begin{table*}[t]
\caption{\label{tab:table1} Scaling of different quantities in the action integral. The first row lists the quantities being scaled. The second row shows the corresponding measuring units needed to effect the scaling. Subscripts $i$ here indicate any of the components $r,\theta,z$. }
\begin{ruledtabular}
\begin{tabular}{ccccccccccccccccccccc}
\textrm{distance}&
\textrm{$v_{i}$}&
\textrm{$\beta_{i}$}&
\textrm{$\gamma$}&
\textrm{$t$}&
\textrm{$f$}&
\textrm{$\dot{t}$}&
\textrm{$\dot{r}$}&
\textrm{$A_{i}$}&
\textrm{$\Phi$}&
\textrm{$E_{i}$}&
\textrm{$H_{i}$}&
\textrm{$B_{i}$}&
\textrm{Energy}&
\textrm{$\mathsf{P}_{i}$\footnote{$\mathsf{P}_{i}$ denotes mechanical momenta. Note that \emph{before} scaling: $\mathsf{p}_{r}=\mathsf{P}_{r}+e_{c}A_{r}$, $\mathsf{p}_{t}=-\textrm{Energy}-e_{c}\Phi$, and $\mathsf{p}_{\theta}/r=\mathsf{P}_{\theta}+e_{c}A_{\theta}$. }}&
\textrm{$\dot{\mathsf{P}}_{i}$}&
\textrm{$\mathsf{p}_{r}$\footnote{$\mathsf{p}_{i}$ denotes canonical momenta. Note that \emph{after} scaling: $\bar{\mathsf{p}}_{r}=\bar{\mathsf{P}}_{r}+\bar{A}_{r}$, $\bar{\mathsf{p}}_{t}=-\bar{\textrm{Energy}}-\bar{\Phi}$, and $\bar{\mathsf{p}}_{\theta}/r=\bar{\mathsf{P}}_{\theta}+\bar{A}_{\theta}$.}}&
\textrm{$\mathsf{p}_{t}$}&
\textrm{$\mathsf{p}_{\theta}$}&
\textrm{$\mathfrak{L}$}&
\textrm{$I$}\\
\colrule
$\lambda$ & $c$ & 1 & 1 & $\frac{\lambda}{c}$ & $\frac{c}{\lambda}$ & $\frac{1}{c}$ & 1 & $\frac{mc}{e_{c}}$ & $\frac{mc^{2}}{e_{c}}$ & $\frac{mc^{2}}{e_{c}\lambda}$ &  $\frac{mc}{e_{c}\lambda\mu_{0}}$ & $\frac{mc}{e_{c}\lambda}$ & $mc^{2}$ & $mc$ & $\frac{mc}{\lambda}$ & $mc$ & $mc^{2}$ & $mc\lambda$ & $mc$ & $mc\lambda$\\
\end{tabular}
\end{ruledtabular}
\end{table*}

\subsection{\label{Hamiltonian} Calculation of $\langle \Delta\gamma^{c} \rangle$ through the Hamiltonian and the equations of motion}

The discussion in Section~\ref{XandY} has indicated, using perturbation theory for a Klystron input cavity with an initially-uniform beam distribution, that $\langle \Delta\gamma^{c} \rangle$ can be approximated by its second-order term average, $\langle \Delta\gamma^{c} \rangle_{k,j}\approx|\alpha_{i}|^{2}\langle \Delta\gamma^{c}_{2} \rangle$, which then allows us to compute the correct detuning parameters ($\bar{X}$ and $\bar{Y}$) for the beam-loaded cavity. In this section, we calculate $\langle \Delta\gamma^{c} \rangle$ by solving the equations of motion for $\gamma^{c}$ and averaging its value over a finite (and relatively small) number of macroparticle beamlets and phase bins, as discussed in Section~\ref{XandY}. We start by deriving the Lagrangian, $\mathfrak{L}$, and Hamiltonian, $\mathfrak{H}$, functions for the field and particle--field interaction, while taking $z$ as the independent variable and working in cylindrical spacial coordinates $(r,\theta,z)$. Taking a Minkowski space with a metric signature $(+,-,-,-)$, a four-vector for position $dx^{\mu}=(c dt,dr,rd\theta,dz)$ and a four-vector for potential $A^{\mu}=(\Phi/c, A_{r},A_{\theta},A_{z})$, where $\Phi$ is the scalar potential and $\bm{A}$ is the vector potential, we can write the action in terms of the integral of relativistic invariants as
\begin{equation}
    I=-\int\limits_\textrm{initial}^\textrm{final} mc\ ds -\int\limits_\textrm{initial}^\textrm{final} e_{c}\ A_{\mu}dx^{\mu}\equiv \int\limits_{z_{i}}^{z_{f}} dz\ \mathfrak{L}, \label{eq:Action1}
\end{equation}
where $ds=\sqrt{dx_{\mu}dx^{\mu}}=\sqrt{c^{2}dt^{2}-dr^{2}-r^{2}d\theta^{2}-dz^{2}}$ and the integration domain is from an initial phase space state to a final one. 

Isolating $dz$ leads to $ds=dz\sqrt{(c\dot{t})^{2}-\dot{r}^{2}-(r\dot{\theta})^{2}-1}$, where here the dot denotes the total derivative in $z$ (e.g.~$\dot{r}=dr/dz$). If we isolate $dt$, we arrive at the relation $ds=cdt\sqrt{1-\beta_{r}^{2}-\beta_{\theta}^{2}-\beta_{z}^{2}}=cdt/\gamma$. Equating these two expressions for $ds$, we obtain the useful relation
\begin{equation}
    \frac{\dot{t}}{\gamma}=\frac{1}{c}\sqrt{(c\dot{t})^{2}-\dot{r}^{2}-(r\dot{\theta})^{2}-1}. \label{eq:keyrelation1}
\end{equation}

Similarly, the term $A_{\mu}dx^{\mu}=\Phi dt-A_{r}dr-rA_{\theta}d\theta-A_{z}dz$ can be reduced by isolating $dz$ to obtain $A_{\mu}dx^{\mu}=dz(\Phi\dot{t}-A_{r}\dot{r}-A_{\theta}r\dot{\theta}-A_{z})$. Equation (\ref{eq:Action1}) thus becomes
\begin{multline}
    I=-mc\int\limits_{z_{i}}^{z_{f}} dz\left[  \sqrt{(c\dot{t})^{2}-\dot{r}^{2}-(r\dot{\theta})^{2}-1} \right. \\
    + \left. \frac{e_{c}}{mc}\left(\Phi\dot{t}-A_{r}\dot{r}-A_{\theta}r\dot{\theta}-A_{z} \right) \right]. \label{eq:I_BeforeScaling}
\end{multline}

We scale the different quantities in (\ref{eq:I_BeforeScaling}) to simplify practical computation, by measuring them using Table~\ref{tab:table1}. The footnotes of Table~\ref{tab:table1} also include connecting formulae, before and after scaling, for mechanical and canonical momenta that will be shortly discussed.  Denoting scaled quantities with an over-bar, (\ref{eq:I_BeforeScaling}) now becomes
\begin{multline}
    \bar{I}=-mc\int\limits_{z_{i}}^{z_{f}} \lambda d\bar{z}\left[  \sqrt{\bar{\dot{t}}^{2}-\bar{\dot{r}}^{2}-(\bar{r}\bar{\dot{\theta}})^{2}-1} \right. \\ + \left.\left(\bar{\Phi}\bar{\dot{t}}-\bar{A}_{r}\bar{\dot{r}}-\bar{A}_{\theta}\bar{r}\bar{\dot{\theta}}-\bar{A}_{z} \right) \right].\label{eq:I_AfterScaling}
\end{multline}

The scaled Lagrangian $\bar{\mathfrak{L}}$ can thus be deduced as
\begin{eqnarray}
    &&\bar{\mathfrak{L}}=\bar{\mathfrak{L}}(\bar{r},\bar{\theta},\bar{t},\bar{\dot{r}},\bar{\dot{\theta}},\bar{\dot{t}};\bar{z}) \nonumber\\
    &&={-}\sqrt{\bar{\dot{t}}^{2}{-}\bar{\dot{r}}^{2}{-}(\bar{r}\bar{\dot{\theta}})^{2}-1} -\bar{\Phi}\bar{\dot{t}}+\bar{A}_{r}\bar{\dot{r}}+\bar{A}_{\theta}\bar{r}\bar{\dot{\theta}}+\bar{A}_{z}.   \label{LLL}
\end{eqnarray}

Using $\bar{\mathfrak{L}}$ we can now define canonical momenta. To avoid confusion with symbols $p$ and $P$ for power, in the following equations we use the symbols $\mathsf{P}$ for mechanical momenta and $\mathsf{p}$ for canonical momenta. The canonical momenta are 
\begin{eqnarray}
    \bar{\mathsf{p}}_r&=&\frac{\partial \bar{\mathfrak{L}}}{\partial \bar{\dot{r}}}=\frac{\bar{\dot{r}}}{\sqrt{\bar{\dot{t}}^{2}-\bar{\dot{r}}^{2}-(\bar{r}\bar{\dot{\theta}})^{2}-1}}+\bar{A_{r}}, \label{eq:p_r} \\
    \bar{\mathsf{p}}_t&=&\frac{\partial \bar{\mathfrak{L}}}{\partial \bar{\dot{t}}}=\frac{-\bar{\dot{t}}}{\sqrt{\bar{\dot{t}}^{2}-\bar{\dot{r}}^{2}-(\bar{r}\bar{\dot{\theta}})^{2}-1}}-\bar{\Phi}, \label{eq:p_t}\\
    \frac{\bar{\mathsf{p}}_{\theta}}{\bar{r}}&=&\frac{\partial \bar{\mathfrak{L}}}{\partial \bar{\dot{\theta}}}=\frac{\bar{r}\bar{\dot{\theta}}}{\sqrt{\bar{\dot{t}}^{2}-\bar{\dot{r}}^{2}-(\bar{r}\bar{\dot{\theta}})^{2}-1}}+\bar{A_{\theta}}, \label{eq:p_theta}
\end{eqnarray}

Note that, after scaling, relation (\ref{eq:keyrelation1}) becomes
\begin{equation}
    \frac{\bar{\dot{t}}}{\bar{\gamma}}=\sqrt{\bar{\dot{t}}^{2}-\bar{\dot{r}}^{2}-(\bar{r}\bar{\dot{\theta}})^{2}-1}, \label{eq:keyrelation1scaled}
\end{equation}
which can be combined with (\ref{eq:p_r})--(\ref{eq:p_theta}) to give the following useful relation, relating the generalized velocities $(\bar{\dot{r}}, \bar{\dot{t}}, \bar{r}\bar{\dot{\theta}})$, the potential fields, $\bar{\gamma}$ and canonical momenta $(\bar{\mathsf{p}_{r}},\bar{\mathsf{p}_{\theta}},\bar{\mathsf{p}_{t}})$ as
\begin{eqnarray}
    \frac{\bar{\dot{t}}}{\bar{\gamma}}&=&\sqrt{\bar{\dot{t}}^{2}-\bar{\dot{r}}^{2}-(\bar{r}\bar{\dot{\theta}})^{2}-1} \nonumber\\
    &=&\frac{1}{\sqrt{  (\bar{\mathsf{p}}_{t}+\bar{\Phi})^{2} {-}(\bar{\mathsf{p}}_{r}-\bar{A}_{r})^{2} {-}(\bar{\mathsf{p}}_{\theta}/\bar{r}-\bar{A}_{\theta})^{2} {-}1  }}. \label{eq:keyRelation2}
\end{eqnarray}

Considering our phase-space canonical variables to be $\bar{\bm{\mathsf{q}}}=(\bar{r},\bar{\theta},\bar{t}), \bar{\bm{\mathsf{p}}}=(\bar{\mathsf{p}_{r}},\bar{\mathsf{p}_{\theta}},\bar{\mathsf{p}_{t}})$, we may now use Legendre transformation \cite{Percival,Goldstein}, $\bar{\mathfrak{H}}=\sum_{i}\bar{\dot{q}}_{i}\bar{p}_{i}-\bar{\mathfrak{L}}$, to find the Hamiltonian as
\begin{equation}
    \bar{\mathfrak{H}}=-\sqrt{ (\bar{\mathsf{p}}_{t}+\bar{\Phi})^{2} {-}(\bar{\mathsf{p}}_{r}-\bar{A}_{r})^{2} {-}(\bar{\mathsf{p}}_{\theta}/\bar{r}-\bar{A}_{\theta})^{2} {-}1}-\bar{A}_{z}. \label{eq:H}
\end{equation}

One of the benefits of the Hamiltonian formalism is its ability to easily reveal symmetries and conserved quantities in the dynamical system, allowing us to identify cyclic (ignorable) variables and to exploit this knowledge to reduce the computational burden as much as possible \cite{Litchenberg,Goldstein}. Indeed, for the problem in hand we take advantage of the symmetry in $\theta$ to reduce the number of the equations of motion to be solved from 6 to 4 or 5 equations, as will be shown below. The conservation of the azimuthal canonical momentum $\mathsf{p}_{\theta}$ underlies this fact (also known as Busch's theorem in \cite{Reiser}). Different field configurations (terms) can be included into or removed from (\ref{eq:H}) in a straightforward manner, before solving for the equations of motion. For brevity, we demonstrate this through two cases with different field configurations (ignoring space charge effects).

In the first case (\emph{case I}) no focusing magnetic field is present (no confinement), where we take $\mathsf{p}_{\theta}=0, A_{\theta}=0, \Phi=0$, with assumed initial conditions $\mathsf{P}_{r_{0}}=\mathsf{P}_{\theta_{0}}=0, \gamma_{0}=1+V_{b}/V_{0}$. In the second case (\emph{case II}) a finite magnetic field is allowed (axial confinement), where we take $\Phi=0$ with the same initial conditions for the first case. It is noted that the strength of $B_{z}$ in practical klystron applications is usually around 1.5 times Brillouin's magnetic flux \cite{Antonsen2002,Antonsen2004,Gilmor}. For case II, it is assumed that $B_{z}$ is uniform in the radial direction. It is also assumed to be zero ($B_{z},A_{\theta}=0$) at the electron gun's position, after which it ramps up linearly in the axial direction to reach its fixed amplitude after a distance $z=l_{0}$ from the gun and stays uniform along the axis thereafter. In this model, $l_{0}$ is chosen to be $(L-g)/6$, where $L$ is the beam pipe length and $g$ is the gap width (see Fig~\ref{fig:inputcavity}).

By substituting (\ref{eq:H}) into the canonical equations of motion, $\dot{\mathsf{q}}_{i}=\partial\mathfrak{h}/\partial\mathsf{p}_{i}, \     \dot{\mathsf{p}}_{i}=-\partial\mathfrak{h}/\partial\mathsf{q}_{i}$, where here $i$ indicates the components $r,\theta$ or $t$, we arrive at the explicit equations of motions for each case, which can be solved numerically. After some algebraic manipulation, we can cast the final equations of motion conveniently in terms of $\gamma, \mathsf{P}_{r},\mathsf{P}_{\theta}$, the fields $E_{r}, E_{z}, B_{z}$ and the variables $r,t,z$ for practical computation, as given below. In addition to the equations of motion, we also include the equation for the imaginary quantity $\breve\gamma$ defined in Subsection~\ref{XandY}, since both $\gamma$ and $\breve{\gamma}$ are needed in (\ref{eq:XandY_FINAL}) for the deduction of $\bar{X}$ and $\bar{Y}$. As expected, it is noted that in both cases, $\mathsf{p}_{\theta}$ is conserved ($\dot{\mathsf{p}}_{\theta}=0$). It is also noted that, since $\Phi=0$, we have $\bar{\gamma}\equiv-\bar{\mathsf{p}}_{t}$.

For case I, we have the equations
\begin{eqnarray}
    \bar{\dot{r}}&=&\frac{\bar{\mathsf{P}}_{r}}{\sqrt{\bar{\gamma}^{2}-\bar{\mathsf{P}}_{r}^{2}-1}} \label{eq:markA} \\
    \bar{\dot{t}}&=&\frac{\bar{\gamma}}{\sqrt{\bar{\gamma}^{2}-\bar{\mathsf{P}}_{r}^{2}-1}} \\
    \bar{\dot{\mathsf{P}}}_{r}&=&\bar{\dot{t}}\ \textrm{Re}\left[\bar{E}_{r}(\bar{r},\bar{t},\bar{z})\right]-\textrm{Re}\left[\bar{B}_{\theta}(\bar{r},\bar{t},\bar{z})\right]\\
    \bar{\dot{\gamma}}&=&\bar{\dot{r}}\ \textrm{Re}\left[\bar{E}_{r}(\bar{r},\bar{t},\bar{z})\right]+\textrm{Re}\left[\bar{E}_{z}(\bar{r},\bar{t},\bar{z})\right] \\
    \bar{\dot{\breve\gamma}}&=&\bar{\dot{r}}\ \textrm{Im}\left[\bar{E}_{r}(\bar{r},\bar{t},\bar{z})\right]+\textrm{Im}\left[\bar{E}_{z}(\bar{r},\bar{t},\bar{z})\right] \label{eq:markB}
\end{eqnarray}
and for case II, the more general of the two cases, we have 
\begin{eqnarray}
    \bar{\dot{r}}&=&\frac{\bar{\mathsf{P}}_{r}}{\sqrt{\bar{\gamma}^{2}-\bar{\mathsf{P}}_{r}^{2}-\bar{\mathsf{P}}_{\theta}^{2}-1}} \label{eq:mark1}\\
    \bar{\dot{t}}&=&\frac{\bar{\gamma}}{\sqrt{\bar{\gamma}^{2}-\bar{\mathsf{P}}_{r}^{2}-\bar{\mathsf{P}}_{\theta}^{2}-1}} \\
     \bar{\dot{\mathsf{P}}}_{r}&=&  \frac{\bar{\mathsf{P}}_{\theta}\left[ \bar{B}_{z}(\bar{r},\bar{z})+\bar{\mathsf{P}}_{r}/\bar{r}\right]}{\sqrt{\bar{\gamma}^{2}-\bar{\mathsf{P}}_{r}^{2}-\bar{\mathsf{P}}_{\theta}^{2}-1}} \\
     &&+\bar{\dot{t}}\ \textrm{Re}\left[\bar{E}_{r}(\bar{r},\bar{t},\bar{z})\right]-\textrm{Re}\left[\bar{B}_{\theta}(\bar{r},\bar{t},\bar{z})\right]\\
     \bar{\dot{\mathsf{P}}}_{\theta}&=&-\bar{\dot{r}}\left[  \frac{\bar{B}_{z}(\bar{r},\bar{z})}{2}+\frac{\bar{\mathsf{P}}_{\theta}}{\bar{r}} \right]\\
     \bar{\dot{\gamma}}&=&\bar{\dot{r}}\ \textrm{Re}\left[\bar{E}_{r}(\bar{r},\bar{t},\bar{z})\right]+\textrm{Re}\left[\bar{E}_{z}(\bar{r},\bar{t},\bar{z})\right] \label{eq:mark2} \\
    \bar{\dot{\breve\gamma}}&=&\bar{\dot{r}}\ \textrm{Im}\left[\bar{E}_{r}(\bar{r},\bar{t},\bar{z})\right]+\textrm{Im}\left[\bar{E}_{z}(\bar{r},\bar{t},\bar{z})\right] \label{eq:mark3}
\end{eqnarray}

Note that the generalized velocity $\bar{r}\bar{\dot{\theta}}$ equation is given by $\bar{\mathsf{P}}_{\theta}/\sqrt{\bar{\gamma}^{2}-\bar{\mathsf{P}}_{r}^{2}-\bar{\mathsf{P}}_{\theta}^{2}-1}$ and can be calculated from the other quantities found in (\ref{eq:mark1})--(\ref{eq:mark2}), but it does not enter explicitly or independently into this simultaneous set of equations for the motion. 
As expected, if the magnetic focusing field is removed from case II, $\bar{\mathsf{P}}_{\theta}$ vanishes and the equations degenerate to those of case I. This is because in such scenario $\bar{A}_{\theta}$ would be zero everywhere, and since $\bar{\mathsf{p}}_{\theta}=\textrm{const}=0=\bar{r}(\bar{\mathsf{P}}_{\theta}+\bar{A}_{\theta})$, with zero initial azimuthal momentum assumed, then $\bar{\mathsf{P}}_{\theta}$ will remain zero. 

\begin{figure}
\includegraphics[width=8.5cm]{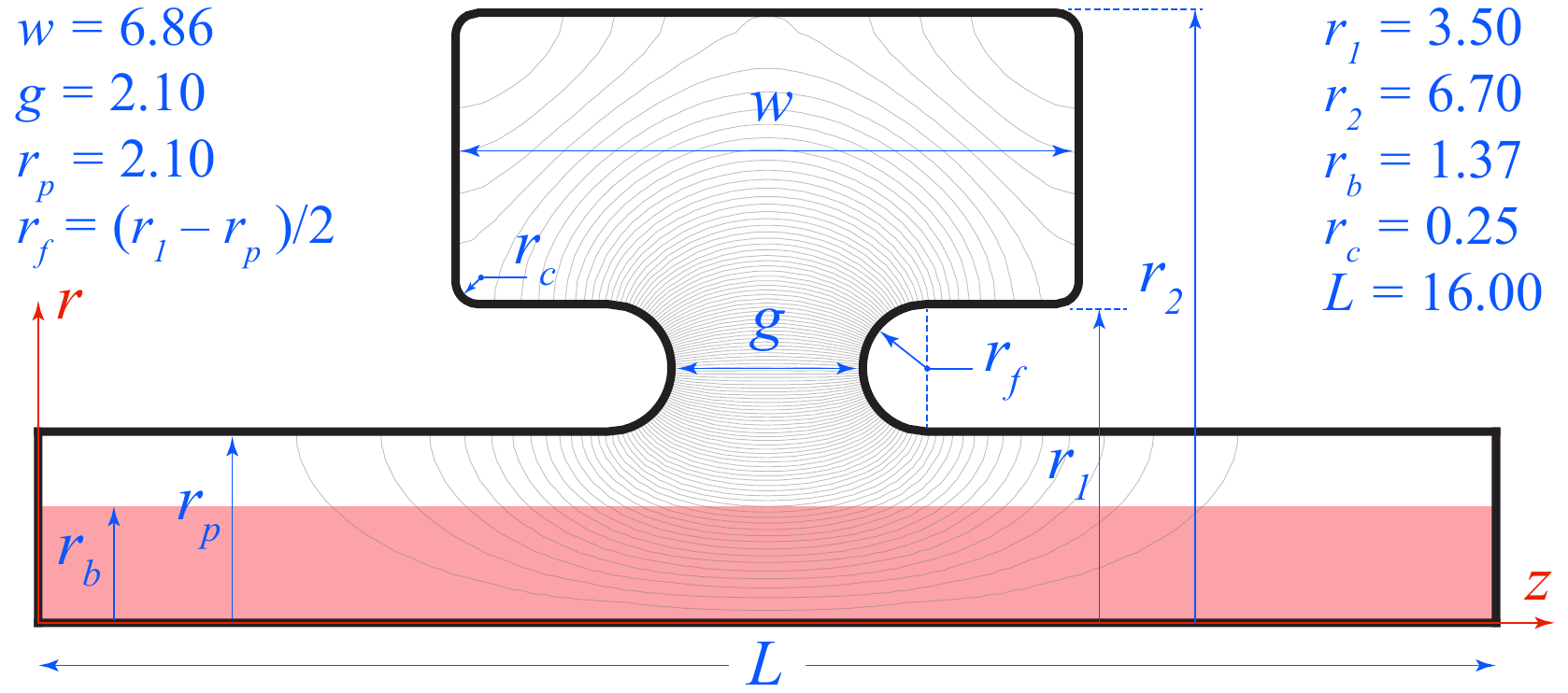}
\caption{\label{fig:example1} Geometry and dimensions of the x-band cavity analyzed in the first example. The figure shows the upper-half of the cavity's cross-section. Electric field streamlines are shown in gray. Dimensions are in mm. }
\end{figure}

\section{\label{Examples} Numerical Examples}

In this section we apply the theory developed in Sections~\ref{Lagrangian} and \ref{InputCavity} to four cavity examples. The theory will predict the beam-loading effects on frequency and coupling coefficient, providing insight into how the detuning can be compensated for and how cavity design can be improved to boost gain in $\langle \Delta \gamma \rangle$. The first example is for an x-band cavity, while the second example uses a c-band cavity that is similar to that used in previous studies \cite{Antonsen2002,Antonsen2004,Antonsen2005} to allow for comparison of beam-loading trends. The third and fourth examples are a c-band cavities with the same frequency and loading conditions as the second example, but with different shapes to demonstrate how shape can influence the $\bar{X}, \bar{Y}$ parameters and the maximum gain achievable by pretuning. For demonstration, we opt to calculate the first example according to the developed equations for case I (no magnetic confinement) and the two other examples according to the equations for case II (with finite magnetic confinement).

The first example comprises the cavity shown in Fig~\ref{fig:example1}, whose cold resonant frequency is $11.424$ GHz. For copper walls this cavity exhibits $Q_{i}=3778$. The cavity has the dimensions shown in the figure and is driven by a beam with $V_{b}=60$~kV and  $I_{b}=10$~A. Calculating $\langle \Delta\gamma^{c} \rangle$ using the equations of motion (\ref{eq:markA})--(\ref{eq:markB}), then using (\ref{eq:XandY_FINAL}), (\ref{eq:FINAL_f_shift}) and (\ref{eq:FINAL_beta}) to extract parameters gives us: $\bar{X}=0.49$, $\bar{Y}=1.40$, a frequency shift of $\Delta f=-2.11$ MHz and new coupling coefficient value of $\beta_{c}=1.49$ for critical coupling.

If we pretune this cavity during design to operate at a small upshift of 2.11 MHz and coupling coefficient of 1.49 (instead of 1), then it will be detuned back to the original frequency when the beam is turned on, and it will then be critically matched as well. With such pretuning we will achieve a gain of $\mathcal{G}_\textrm{max}=1.37$ (equivalent to 1.4~dB) compared to the untuned case, according to (\ref{eq:Gain}).

\begin{figure}
\includegraphics[width=8.5cm]{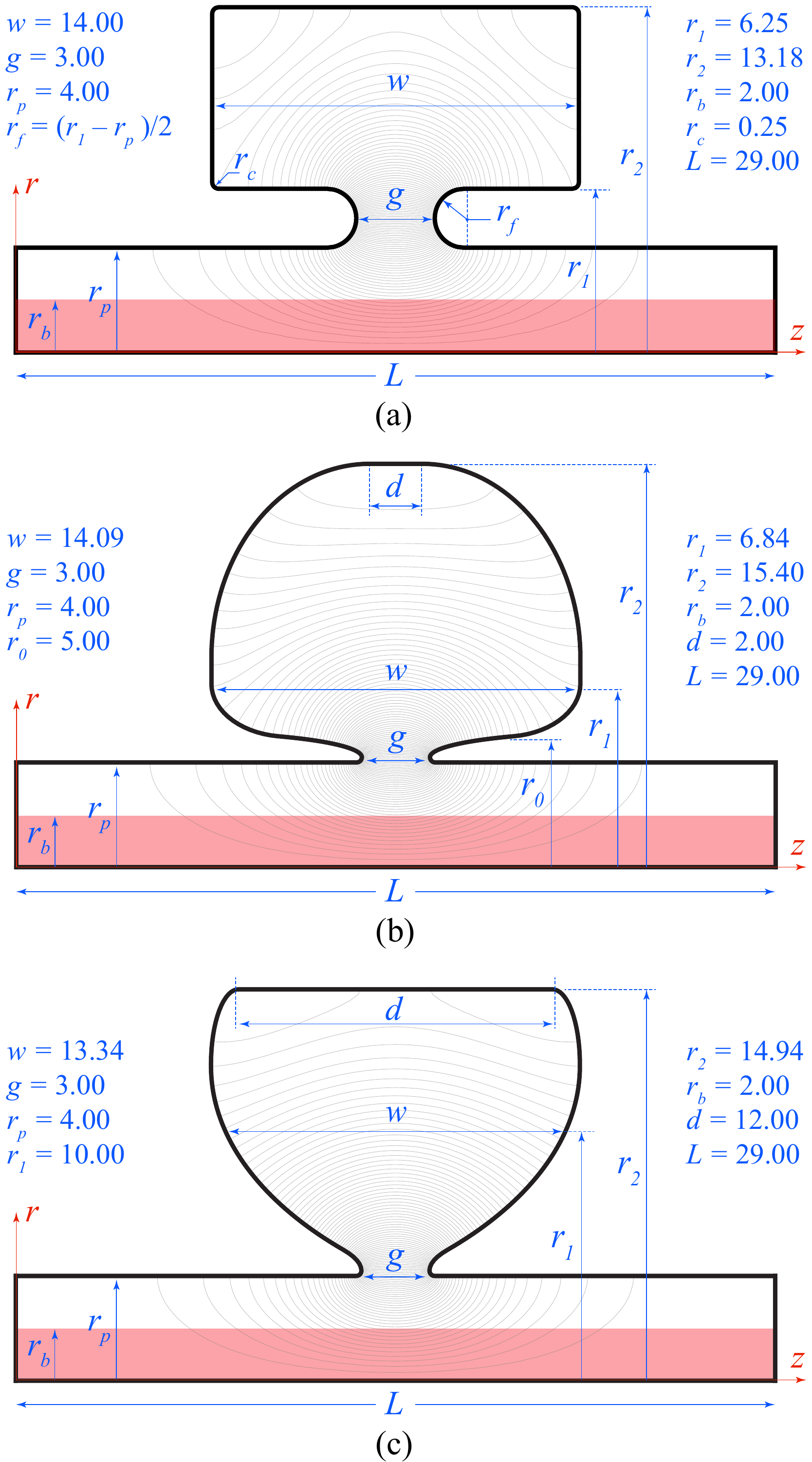}
\caption{\label{fig:example2} Geometries and dimensions of the c-band cavities analyzed in the second (a), third (b) and fourth (c) examples. The figures show the upper-halves of the cavity cross-sections. The cavity in (a) is similar to the one studied in \cite{Antonsen2002,Antonsen2004,Antonsen2005}. Profiles of the ``mushroom" and ``inverted-mushroom" cavities in (b) and (c) are drawn with piecewise-elliptic functions, with guiding dimensions shown (to-scale). The cavity in (b) gives highest maximum gain, $\mathcal{G}_\textrm{max}$, upon pretuning compared to the one in (a) and (c).  Electric field streamlines are shown in gray. Dimensions are in mm. }
\end{figure}

\begin{table}[t]
\caption{\label{tab:table2}
Instances of beam-loading experienced by the c-band cavity shown in Fig~\ref{fig:example2} under fixed perveance of 0.375 $\mu$Pervs. The columns show the detuning predictions by the presented theory as well as the frequency shift predictions reported in \cite{Antonsen2002}. The latter is shown in brackets.}
\begin{ruledtabular}
\begin{tabular}{ccccccc}
 $V_{b}$ (kV)  & $I_{b}$ (A) & $B_{z}$ (T) & $\Delta f$ (MHz) & $\beta_{c} $ & $\bar{X}$ & $\bar{Y}$  \\
\hline
40 & 3 & 1.3 & $-0.46$ ($-0.2$) & 1.22 & 0.22 & 0.91 \\
150 & 21.8 & 1.3 & $-0.41$ ($-1.2$) & 1.15 & 0.15 & 0.81 \\
250 & 46.9 & 1.3 & $-0.36$ ($-1.4$) & 1.09 & 0.09 & 0.72 \\
350 & 77.6 & 2.6 & $-0.32$ ($-1.2$) & 1.07 & 0.07 & 0.64 \\
450 & 113.0 & 2.6 & $-0.29$ ($-1.2$) & 1.06 & 0.06 & 0.58 \\
\end{tabular}
\end{ruledtabular}
\end{table}

For the second example we consider a cavity of a similar shape, but tuned to the frequency 5.08764 GHz [dimensions are shown in Fig~\ref{fig:example2}(a)]. This cavity structure, which is almost identical to that studied in \cite{Antonsen2002,Antonsen2004,Antonsen2005} except for small fillet rounding around the gap corners, is chosen to facilitate the comparison of beam-loading trends as predicted by our theory and those studies. Using copper walls for this example, the cold quality factor is $Q_{i}=5017$. We analyze this example using the equations of case II, (\ref{eq:mark1})--(\ref{eq:mark3}), to measure the effect of perveance, beam voltage and beam current on beam loading, in a manner similar to \cite{Antonsen2002}. It was reported in \cite{Antonsen2002} that changing the beam voltage and current separately, while having the same perveance, will not affect beam-loading considerably and the frequency shift will be negligible. If the voltage is held fixed, however, then the frequency was reported to decrease almost linearly with the perveance (or current). These trends seem to agree with the results from the current analysis, under similar conditions of magnetic flux confinement. Table~\ref{tab:table2} shows the behaviour under fixed perveance, while Table~\ref{tab:table3} shows the behavior under fixed voltage. Calculations from the presented theory and from \cite{Antonsen2002} are compared in the two tables. It is noted that, whilst the two sets of results agree on the trend of behavior, the presented theory seems to predict a shift that is relatively less acute than that predicted by \cite{Antonsen2002}. 

\begin{table}[t]
\caption{\label{tab:table3}
Detuning experienced at different instances of beam-loading for the c-band cavity shown in Fig~\ref{fig:example2}(a) and under fixed beam voltage $V_{b}=240$ kV. The rows show the detuning predictions by the present theory as well as the frequency shift predictions reported in \cite{Antonsen2002} (within reading accuracy of curves given in \cite{Antonsen2002}). The last column gives the approximately linear trends of detuning for the present theory compared to the PIC simulations in \cite{Antonsen2002} and the L-5782 weather radar klystron cited in \cite{Antonsen2002}.}
\begin{ruledtabular}
\begin{tabular}{lccccc|c}
  $I_{b}$ (A) & 118.0 & 164.5 & 187.5 & 211.4 & 235.2 & slope \\
\hline
$\Delta f$:                           & \scriptsize{MHz}    & \scriptsize{MHz}    & \scriptsize{MHz}    & \scriptsize{MHz}    & \scriptsize{MHz}    & \scriptsize{MHz/$\mu$Pervs}\\
(1) This Theory                 & $-1.0$ & $-1.4$ & $-1.6$ & $-1.8$ & $-2.0$ & circa $-1$  \\
(2) L-5782 rdr~\cite{Antonsen2002} & $-$ & $-$ & $-$ & $-$ & $-$ & circa $-2$ \\
(3) PIC sim.~\cite{Antonsen2002} & $-3.4$ & $-5.0$ & $-6.4$ & $-7.2$ & $-8.2$ & circa $-4.9$ \\
 
\hline
This Theory, $\beta_{c}$ & $1.24$ & $1.36$ & $1.38$ & $1.43$ & $1.48$ &  \\
This Theory, $\bar{X}$  & $0.24$ & $0.36$ & $0.38$ & $0.43$ & $0.48$ &  \\
This Theory, $\bar{Y}$  & $1.95$ & $2.72$ & $3.10$ & $3.50$ & $3.90$ &  \\
\end{tabular}
\end{ruledtabular}
\end{table}

\begin{figure}
\includegraphics[width=7.5cm]{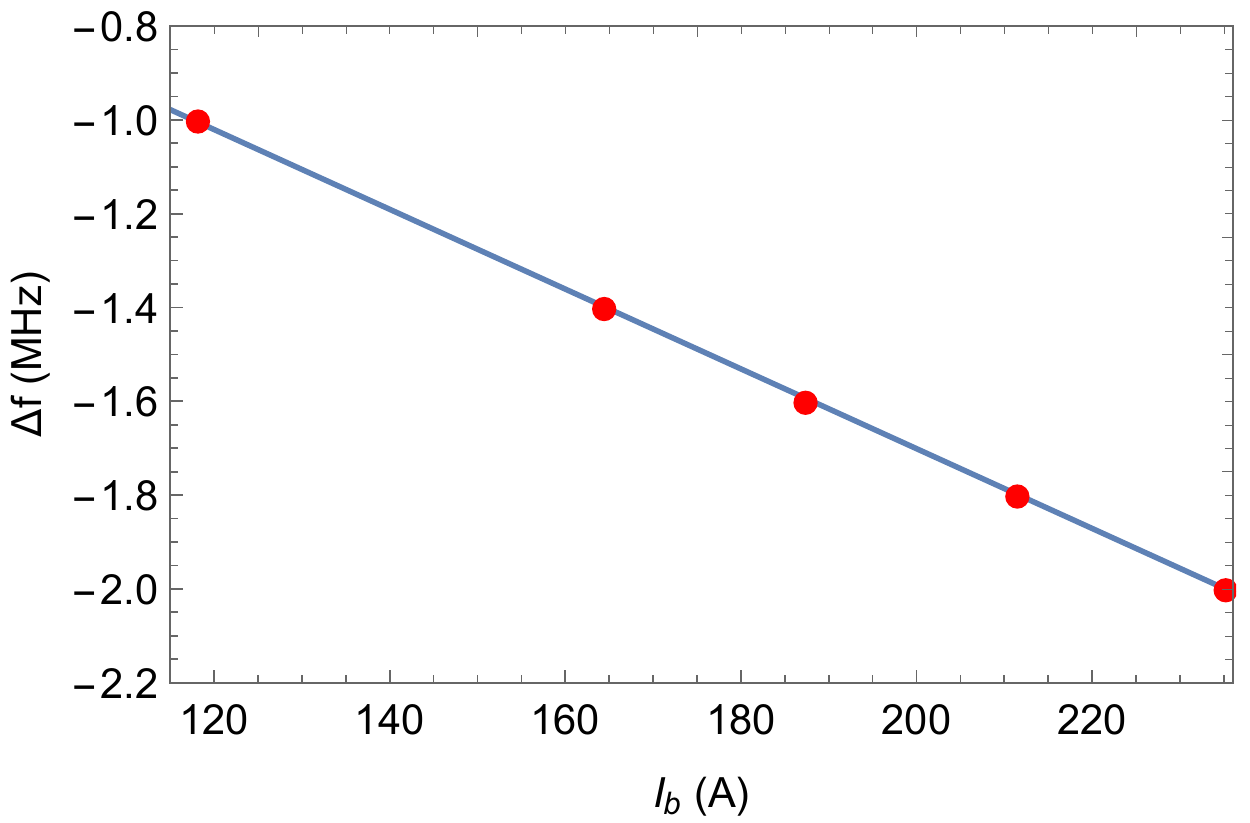}
\caption{\label{fig:figure5} Frequency shift experienced by the c-band cavity shown in Fig~\ref{fig:example2}(a) under fixed beam voltage $V_{b}=240$ kV and varying perveance (current), as predicted by the presented theory, with sample current values chosen to match those presented by \cite{Antonsen2002}. The almost linear trend is in agreement with general trend reported in \cite{Antonsen2002}, but with a lower slope. See Table~\ref{tab:table3} and text for details.}
\end{figure}

Fig~\ref{fig:figure5} shows the approximately linear behaviour of $\Delta f$ with beam current, at a fixed voltage of 240 kV and 1.3 Tesla confinement. Our theory predicts a trend where frequency is shifted by approximately $-1$ MHz per $\mu$Perv. This should be compared by the prediction of $-4.9$ MHz/$\mu$Perv in the PIC simulations reported in \cite{Antonsen2002} and with the estimation rate of $-2$ MHz/$\mu$Perv given for the L-5782 weather radar klystron, cited in \cite{Antonsen2002}. 

Let us consider this cavity at the loading point of $V_{b}=240$~kV $I_{b}=235.2$~A (see Table~\ref{tab:table3}), and pretune it to operate at a small upshift of 2.0 MHz and coupling coefficient of 1.48, then it will be detuned back to the original frequency when the beam is turned on, and it will then be critically matched as well. With such pretuning we will achieve a gain of $\mathcal{G}_\textrm{max}=3.61$ (equivalent to 5.6~dB) compared to the untuned case, according to (\ref{eq:Gain}).

Finally, consider the c-band structures shown in Fig.~\ref{fig:example2}(b) and \ref{fig:example2}(c), as two cavities that are designed to operate at the same frequency as the second example above (5.08764~GHz), but with a different, piecewise-elliptical profile shapes.  Analysing these two cavities using the same equations, (\ref{eq:mark1})--(\ref{eq:mark3}), and under same beam-loading point ($V_{b}=240$~kV $I_{b}=235.2$~A) taken for the previous example, we obtain for the mushroom-like cross-section [see Fig.~\ref{fig:example2}(b)]: $\bar{X}=0.75$, $\bar{Y}=6.26$, $\Delta f=-2.43$~MHz and $\beta_{c}=1.75$, for copper walls with a cold $Q_{i}$ of $6555$. For the inverted-mushroom profile [see Fig.~\ref{fig:example2}(c)] we obtain: $\bar{X}=0.61$, $\bar{Y}=5.09$, $\Delta f=-2.01$~MHz and $\beta_{c}=1.61$, with a cold $Q_{i}$ of $6456$. This results in the former having $\mathcal{G}_\textrm{max}=6.69$ (equivalent to 8.3~dB) and the latter having $\mathcal{G}_\textrm{max}=5.09$ (equivalent to 7.1~dB), according to (\ref{eq:Gain}). This indicates that the mushroom-like cavity in Fig.~\ref{fig:example2}(b) has the highest $\mathcal{G}_\textrm{max}$ value of the three cavities in Fig.~\ref{fig:example2}, giving almost twice the gain-ceiling obtained by using Fig.~\ref{fig:example2}(a); an advantage for using the shape in Fig.~\ref{fig:example2}(b) in practice. This highlights how the $\bar{X}, \bar{Y}$ parameters can inform shape optimization to maximize $\mathcal{G}_\textrm{max}$ during cavity design.

\section{\label{Conclusions} Conclusions}
A variational theory for beam-loaded cavities has been presented, where the beam--field interaction is formulated as a dynamical problem. The general theory, which is applicable to an arbitrary cavity setup, is applied to the important class of cylindrical structures exemplified by a klystron input cavity, to extract the laws that govern detuning under beam loading. Two important parameters, $\bar{X}, \bar{Y}$, were extracted to characterize the cavity under beam loading. With the help of these parameters, the cavity detuning behavior is predicted and a pretuning procedure is proposed to bring the cavity back to nominal resonance and to maximize gain in $\langle \Delta \gamma \rangle$. It is shown that the detuning parameters $\bar{X}, \bar{Y}$ can be also used to optimize the cavity's shape during the design process, futher maximizing the gain achieved by pretuning.  The presented theory uses Hamiltonian formalism to express the underlying physics and computations in a relatively straightforward manner. The powerful methods found in Hamiltonian dynamics and infinitesimal canonical transformations (e.g.~Lie transformations) are now directly accessible for the beam loading problem.

\begin{acknowledgments}
The authors would like to thank M.~Shummail, of SLAC, Stanford University, for several interesting discussions in relation to this research. This work was supported
by the US Department of Energy (contract number DE-AC02-76SF00515) and by the US National Science Foundation (grant number 1734015).

\end{acknowledgments}

\appendix

\section{\label{Appx1}Derivation of some preliminary parameters}
For the intrinsic quality factor $Q_{i}$ we can see that (\ref{eq:Qi}) is arrived at by writing
\begin{eqnarray}
    Q_{i}&=&\omega_{i}\frac{U_{i}}{P_{l_{i}}}= \omega_{i}\frac{\int_{V}dV\left( \frac{\mu}{4}|\bm{H}|^{2}+\frac{\epsilon}{4}|\bm{E}|^{2} \right)}{\text{Re} \left[\int_{\partial V}da \ \frac{1}{2}(\bm{E}\times\bm{H}^{*})\cdot\hat{\bm{n}}\right]} \nonumber \\
    &\cong&\omega_{i}\frac{|\alpha_{i}|^{2}\int_{V}dV\left( \frac{\mu}{4}|\bm{h}_{i}|^{2}+\frac{\epsilon}{4}|\bm{e}_{i}|^{2} \right)}{|\alpha_{i}|^{2}\text{Re} (1-i)\frac{\xi}{2}\int_{\partial V}da\ h_{i}^{2}} \nonumber \\
    &=&\omega_{i}\frac{\mu\int_{V}dV h_{i}^{2}}{\xi\int_{\partial V}da\ h_{i}^{2}}=-\omega_{i}\frac{\epsilon\int_{V}dV e_{i}^{2}}{\xi\int_{\partial V}da\ h_{i}^{2}}= \omega_{i}\frac{u_{i}}{p_{i}}. \nonumber
\end{eqnarray}

The relation between the scaled energy terms in (\ref{eq:Ui_uiANDPl_pi}) is readily obtainable by
\begin{eqnarray}
U_{i}&=&\frac{1}{2}\int\limits_{V}dV\mu|\bm{H}|^{2}=\frac{1}{2}|\alpha_{i}|^{2}\int\limits_{V}dV\mu h_{i}^{2} \nonumber \\
&=&\frac{1}{2}\int\limits_{V}dV\epsilon|\bm{E}|^{2}=\frac{e^{-i\pi}}{2}|\alpha_{i}|^{2}\int\limits_{V}dV\epsilon e_{i}^{2}  \nonumber \\
&=&-\frac{1}{2}|\alpha_{i}|^{2}\int\limits_{V}dV\epsilon e_{i}^{2}=|\alpha_{i}|^{2}u_{i}, \nonumber
\end{eqnarray}
and, similarly, $ P_{l_{i}}=|\alpha_{i}|^{2}p_{i}$.

The external quality factor $Q_{e}$ equation (\ref{eq:QeTempEq}) is reached also by writing 
\begin{eqnarray}
    Q_{e}&=&\omega_{i}\frac{U_{e}}{P_{l_{i}}}= \frac{\omega_{i}}{2}\frac{\int_{V}dV\left( \mu|\bm{H}|^{2}+\epsilon|\bm{E}|^{2} \right)}{\int_{p}da \ (\bm{E}_{e}\times\bm{H}_{e}^{*})\cdot\hat{\bm{n}}} \nonumber \\
    &=&\frac{\omega_{i}}{2}e^{-2i(\phi_{H}-\phi_{\text{em}})}\frac{\int_{V}dV\left( \mu H^{2}-\epsilon E^{2} \right)}{\int_{p}da \ (\bm{E}_{e}\times\bm{H}_{e})\cdot\hat{\bm{n}}} \nonumber \\
     &=&\frac{-\omega_{i}e^{-2i\Delta\phi}\int_{V}dV \epsilon E^{2}}{\int_{p}da \ (\bm{E}_{e}\times\bm{H}_{e})\cdot\hat{\bm{n}}} {\cong}\frac{-\omega_{i}e^{-2i\Delta\phi}\alpha^{2}_{i}\int_{V}dV \epsilon e_{i}^{2}}{a^{2}_{e}\int_{p}da \ (\bm{e}_{p}\times\bm{h}_{p})\cdot\hat{\bm{n}}} \nonumber \\    
    &=&\frac{\omega_{i}e^{-2i\Delta\phi}\int_{V}dV \mu H^{2}}{\int_{p}da \ (\bm{E}_{e}\times\bm{H}_{e})\cdot\hat{\bm{n}}} {\cong}\frac{\omega_{i}e^{-2i\Delta\phi}\alpha^{2}_{i}\int_{V}dV \mu h_{i}^{2}}{a^{2}_{e}\int_{p}da \ (\bm{e}_{p}\times\bm{h}_{p})\cdot\hat{\bm{n}}}. \nonumber
\end{eqnarray}

The emitted power, $P_{e}$, can be written as
\begin{equation}
P_{e}=\frac{1}{2}\int\limits_{p} da\ \bm{E}_{e}{\times}\bm{H}^{*}_{e}\cdot \hat{\bm{n}} =\frac{|a_{e}|^{2}}{2}{\int\limits_{p}}da\ \bm{e}_{p}{\times}\bm{h}_{p}\cdot\hat{\bm{n}} \label{eq:P_initial}
\end{equation}

The term $\int_{p}da\ \bm{e}_{p}{\times}\bm{h}_{p}\cdot\hat{\bm{n}}$ in (\ref{eq:P_initial}) may be extracted from  (\ref{eq:Qefinal}) and (\ref{eq:ui}) as
\begin{eqnarray}
\int_{p}da \ (\bm{e}_{p}\times\bm{h}_{p})\cdot\hat{\bm{n}}&=&\frac{-\omega_{i}\alpha^{2}_{i}}{a^{2}_{e}Q_{e}}\int\limits_{V}dV \epsilon e_{i}^{2}\nonumber\\
&=&\frac{2\omega_{i}\alpha^{2}_{i}u_{i}}{a^{2}_{e}Q_{e}}=\frac{2\omega_{i}|\alpha_{i}|^{2}u_{i}}{|a_{e}|^{2}Q_{e}}. \label{eq:Integral_exh}
\end{eqnarray}

Equation (\ref{eq:Integral_exh}) is now reused to find expressions for the power inputted through the port, $P_{\text{in}}$ given in (\ref{eq:Pin}), and consequently the ratio $|a^{+}|/|a_{e}|$ given in (\ref{eq:useful1}), as follows
\begin{eqnarray}
P_{\text{in}}&=&-\frac{1}{2}\int\limits_{p}da\  \bm{E}^{+}\times\bm{H}^{+*}\cdot\hat{\bm{n}}=\frac{|a^{+}|^{2}}{2}\int\limits_{p}da\  \bm{e}_{p}\times\bm{h}_{p}\cdot\hat{\bm{n}}  \nonumber \\
&=&\frac{|a^{+}|^{2}|\alpha_{i}|^{2}}{|a_{e}|^{2}}\frac{\omega_{i}u_{i}}{Q_{e}}=e^{-2i\phi^{+}} \frac{a^{+2}\alpha_{i}^{2}}{a_{e}^{2}}\frac{\omega_{i}u_{i}}{Q_{e}},\nonumber 
\end{eqnarray}
\begin{equation}
    \frac{|a^{+}|}{|a_{e}|}=\frac{1}{|\alpha_{i}|}\sqrt{\frac{P_{\text{in}}Q_{e}}{\omega_{i}u_{i}}}=\frac{e^{i\phi_{\alpha_{i}}}}{\alpha_{i}}\sqrt{\frac{P_{\text{in}}Q_{e}}{\omega_{i}u_{i}}}. \nonumber
\end{equation}


\section{\label{Appx2}Consistency proof for the proposed cavity Lagrangian $\mathcal{L}$ in equation (\ref{eq:L})}

The cavity's driven Lagrangian $\mathcal{L}$  was given in (\ref{eq:L}) as
\begin{eqnarray}
\mathcal{L}&=&\frac{1}{2}\int\limits_{V}dV\left( \epsilon E^{2}+\mu H^{2} \right)+\frac{i}{2\omega}\int\limits_{p}da\ \left( \bm{E}_{e}\times\bm{H}_{e} \right)\cdot \hat{\bm{n}} \nonumber \\
&& +\frac{1+i}{2\omega}\xi\int\limits_{\partial V}da\ \bm{H}^{2} {+}\frac{2i}{\omega}\int\limits_{p}da\ \left( \bm{E}_{e}\times\bm{H}^{+} \right)\cdot \hat{\bm{n}}. \label{Appx:L}
\end{eqnarray}

Consider for a moment the first (volume integral) term in (\ref{Appx:L}), which we shall call $\mathcal{L}_{1}$, and let us now use (\ref{eq:HAsMainVariable}) to work in terms of $\bm{H}$ as the main variable, then we can write
\begin{eqnarray}
\mathcal{L}_{1}&=&\frac{1}{2}\int\limits_{V}dV\left( \epsilon E^{2}+\mu H^{2} \right) \nonumber \\
&=&\frac{1}{2}\int\limits_{V}dV\left[ \epsilon \left( \frac{\nabla\times \bm{H}-\bm{J}}{-i\omega\epsilon} \right)^{2}+\mu H^{2} \right] \nonumber \\
\delta\mathcal{L}_{1}&=&\int\limits_{V}dV\left[ \frac{-1}{\epsilon\omega^{2}} (\nabla\times\delta\bm{H})\cdot( \nabla\times \bm{H}{-}\bm{J}){+}\mu \delta \bm{H}{\cdot}\bm{H} \right]. \nonumber
\end{eqnarray}

Utilizing the vector identity $\nabla\cdot\left[ \delta\bm{H}\times(\nabla\times\bm{H}-\bm{J}) \right]=(\nabla\times\delta\bm{H})\cdot( \nabla\times \bm{H}-\bm{J})-\delta\bm{H}\cdot\nabla\times(\nabla\times\bm{H}-\bm{J})$ and using Divergence theorem now gives
\begin{eqnarray}
\delta\mathcal{L}_{1}&=&\frac{-1}{\epsilon\omega^{2}}\int\limits_{V}dV \delta\bm{H}\cdot\nabla\times(\nabla\times\bm{H}-\bm{J}){+}\int\limits_{V}dV\mu \delta \bm{H}{\cdot}\bm{H} \nonumber\\
&& + \frac{-1}{\epsilon\omega^{2}}\oint\limits_{\partial V}da\ \hat{\bm{n}}\cdot\delta\bm{H}\times(\nabla\times\bm{H}-\bm{J}) \nonumber \\
&=&\int\limits_{V}dV \delta\bm{H}{\cdot}\left(\frac{i}{\omega}\nabla{\times}\bm{E}{+}\mu \bm{H}\right) {+}\frac{i}{\omega}\oint\limits_{\partial V} da\ \delta\bm{H}{\cdot}\hat{\bm{n}}\times\bm{E} \nonumber\\
&=&\int\limits_{V}dV \delta\bm{H}{\cdot}\left(\frac{i}{\omega}\nabla{\times}\bm{E}{+}\mu \bm{H}\right) {+} \frac{i}{\omega}\int\limits_{\textrm{wall}} da\ \delta\hat{\bm{H}}{\cdot}\hat{\bm{n}}\times\bm{E}\nonumber\\
&& +\frac{i}{\omega}\int\limits_{p} da\ \left.\delta\bm{H}\right|_{p}{\cdot}\left.\hat{\bm{n}}\times\bm{E}\right|_{p}, \label{eq:Appx:stepPP}
\end{eqnarray}
where we have decomposed the surface $\partial V$ into the part covering the walls and the part covering the port. If we now add the rest of the terms from (\ref{Appx:L}) and force $\delta\mathcal{L}$ to vanish for an arbitrary $\delta\bm{H}$ variation, it is easy to see that the volume and surface integrals must vanish for arbitrary $\delta\bm{H}$, forcing their integrands to satisfy the following relations
\begin{eqnarray}
\nabla\times\bm{E}&=&i\omega\mu\bm{H}, \\
\left.\bm{E}\right|_{p}&=&\bm{E}_{e}, \ \ \left. \bm{H}\right|_{p}=\bm{H}_{e}+2\bm{H}^{+} \ \ (\textrm{on port}), \\
\hat{\bm{n}}\times\bm{E}&=&(1-i)\xi\bm{H} \ \ (\textrm{on walls}),
\end{eqnarray}
which are exactly Maxwell's second equation and the boundary conditions expected at the walls and the port, as required.


\section{\label{Appx3}Derivation of frequency shift due to cavity wall losses using the undriven Lagrangian $\mathcal{L}$}
Consider the undriven Lagrangian, including only the effect of cavity wall-loss and denoting the corresponding frequency by $\omega'_{i}$,
\begin{eqnarray}
\mathcal{L}&=&\frac{1}{2}\int\limits_{V}dV\left( \epsilon E^{2}+\mu H^{2} \right) +\frac{1+i}{2\omega'_{i}}\xi\int\limits_{\partial V}da\ \bm{H}^{2}, \label{Appx:Lundriven}
\end{eqnarray}
and let us apply the modal approximation of (\ref{eq:EandH}), as well as (\ref{eq:HAsMainVariable}) with $\bm{J}$ set to zero, to give
\begin{eqnarray}
\mathcal{L}&=&\frac{1}{2}\int\limits_{V}dV\left( \epsilon E^{2}+\mu H^{2} \right) +\frac{1+i}{2\omega'_{i}}\xi\int\limits_{\partial V}da\ \bm{H}^{2} \nonumber\\
&=&\frac{1}{2}\int\limits_{V}dV\left[ \epsilon \frac{(\nabla\times\bm{H})^{2}}{-\omega_{i}^{'2}\epsilon^{2}}+\mu H^{2} \right] +\frac{1+i}{2\omega'_{i}}\xi\int\limits_{\partial V}da\ \bm{H}^{2} \nonumber\\
&=&\frac{1}{2}\int\limits_{V}dV\left[ \frac{-\alpha_{i}^{2}}{\omega_{i}^{'2}\epsilon}(\nabla{\times}\bm{h}_{i})^{2}{+}\mu\alpha_{i}^{2} h_{i}^{2} \right] {+}\frac{1+i}{2\omega'_{i}}\alpha_{i}^{2}\xi\int\limits_{\partial V}da h_{i}^{2} \nonumber \\
&=&\frac{1}{2}\int\limits_{V}dV \alpha_{i}^{2}\mu h_{i}^{2}\left[ 1{-}\left(\frac{\omega_{i}}{\omega'_{i}}\right)^{2} \right] {+}\frac{1+i}{2\omega'_{i}}\alpha_{i}^{2}\xi\int\limits_{\partial V}da h_{i}^{2}, \label{Appx:Anotherstep} 
\end{eqnarray}
where in the last step we exploited the relation given in (\ref{eq:ui}). If we use (\ref{eq:useful3}) in (\ref{Appx:Anotherstep}) to convert the surface integral into a volume integral, we can write
\begin{eqnarray}
\mathcal{L}&=&\frac{1}{2}\int\limits_{V}dV \alpha_{i}^{2}\mu h_{i}^{2}\left[ 1-\left(\frac{\omega_{i}}{\omega'_{i}}\right)^{2}  +\frac{(1+i)}{Q_{i}}\frac{\omega_{i}}{\omega'_{i}}\right] \nonumber \\
&=&\frac{1}{2}\int\limits_{V}dV \alpha_{i}^{2}\mu h_{i}^{2} \frac{\omega_{i}}{\omega'_{i}} \left[ \frac{\omega'_{i}}{\omega_{i}}-\frac{\omega_{i}}{\omega'_{i}}  +\frac{(1+i)}{Q_{i}}\right]. \label{Appx:final}
\end{eqnarray}

It is now easy to see that, by invoking the stationarity of $\mathcal{L}$, equation (\ref{Appx:final}) will, indeed, lead to the expected result (as given in classical treatments, e.g. \cite{slaterbook}) for the complex frequency shift due to the finite-conductivity of the cavity walls,
\begin{equation}
   \left(\frac{\omega'_{i}}{\omega_{i}}-\frac{\omega_{i}}{\omega'_{i}}\right)=-\frac{1+i}{Q_{i}}, \label{eq:FreqShiftInitial_duplicate}
\end{equation}
which can be simplified for high-$Q_{i}$ cavities as
\begin{equation}
    \frac{2\Delta\omega_{i}}{\omega'_{i}}\cong -\frac{1+i}{Q_{i}}.  \label{eq:FreqShift}
\end{equation}

Although the shifted frequency in (\ref{eq:FreqShift}) is complex in general and may be written as $\omega'_{i}=\omega'_{i,r}+i\omega'_{i,x}$, with the imaginary part representing damping effects, it can be readily shown that the same form of equation (\ref{eq:FreqShift}) is found for the real frequency shift ($\omega_{i}\rightarrow\omega'_{i,r}$). Indeed, one can write
\begin{equation}
-\frac{1+i}{Q_{i}}\cong\frac{2\Delta\omega_{i}}{\omega'_{i}}=\frac{2(\omega'_{i,r}-\omega_{i}+i\omega'_{i,x})}{\omega'_{i,r}+i\omega'_{i,x}}=\frac{2(\Delta\omega_{i,r}+i\omega'_{i,x})}{\omega'_{i,r}+i\omega'_{i,x}}, \nonumber
\end{equation}
whose real and imaginary parts can be separated and solved simultaneously to give
\begin{eqnarray}
\omega'_{i,x}&\cong&\frac{-\omega'_{i,r}}{1+2Q_{i}} \label{eq:omega_x}, \\ \frac{2\Delta\omega_{i,r}}{\omega'_{i,r}}&\cong&\left(\frac{\omega'_{i,r}}{\omega_{i}}-\frac{\omega_{i}}{\omega'_{i,r}}\right)=-\frac{1}{Q_{i}} \label{eq:delta_duplicate}, \\
\Rightarrow \omega'_{i,r}&\cong& \omega_{i}\left( 1-\frac{1}{2Q_{i}}\right). \label{eq:omega_r_duplicate}
\end{eqnarray}

As expected \cite{slaterbook}, the frequency correction for the eigenmode is small for high-$Q$ cavities. The small imaginary part's relative magnitude is of the same order-of-magnitude as the frequency correction; $\mathcal{O}(1/Q_{i})$.  Note that, for the type of time-harmonic variational formulation used in Section~\ref{Lagrangian}, the frequency enters the driven Lagrangian formalism as the real frequency \cite{schwinger}. Therefore, in Section~\ref{Lagrangian} we use the notation $2\Delta\omega_{i}/\omega'_{i}$, dropping the $r$ subscript in $\omega'_{i}$ for simplicity, with the understanding that we are dealing with real frequencies.

\vfill


\bibliography{myrefs}

\end{document}